\documentclass[11pt]{article}
\usepackage{fullpage}
\usepackage{setspace}
\usepackage{url}
\usepackage{xcolor}
\usepackage{parskip}
\usepackage{float}
\usepackage[section]{placeins}
\usepackage[square,numbers]{natbib}
\usepackage{graphicx}
\usepackage[labelformat=simple]{subcaption}
\usepackage{amsmath}
\usepackage{algorithm2e}

\begin{document}
\doublespacing

% Title of paper
\title{Assessing effect heterogeneity of a randomized treatment using conditional inference trees}

% List of authors, with corresponding author marked by asterisk
\author{ASHWINI VENKATASUBRAMANIAM$^\mathsection$, BRANDON KOCH$^\ddagger$,  LAUREN ERICKSON$^\mathparagraph$,\\ SIMONE FRENCH$^\star$, DAVID VOCK$^\dagger$ and JULIAN WOLFSON$^\dagger$\\[4pt]
% Author addresses
\textit{$^\mathsection$The Alan Turing Institute, British Library, London, UK}\\
\textit{$^\ddagger$School of Community Health Sciences, University of Nevada, Reno, USA}\\
\textit{$^\mathparagraph$HealthPartners Institute for Education and Research, Minneapolis, Minnesota, USA}\\
\textit{$^\star$Division of Epidemiology and Community Health, School of Public Health,} \\
\textit{University of Minnesota, Minneapolis, USA}\\
\textit{$^\dagger$Division of Biostatistics, School of Public Health, University of Minnesota, Minneapolis, USA}\\
% E-mail address for correspondence
}

% Running headers of paper:
\markboth
% First field is the short list of authors
{A. Venkatasubramaniam, B. Koch, L. Erickson, S. French, D. Vock and J. Wolfson}
% Second field is the short title of the paper
{Treatment effect heterogeneity using conditional inference trees}

\maketitle
% Add a footnote for the corresponding author if one has been
% identified in the author list
%\footnotetext{To whom correspondence should be addressed. Email: {julianw@umn.edu}}

\begin{abstract}
{Treatment effect heterogeneity occurs when individual characteristics influence the effect of a treatment. We propose a novel approach that combines prognostic score matching and conditional inference trees to characterize effect heterogeneity of a randomized binary treatment. One key feature that distinguishes our method from alternative approaches is that it controls the Type I error rate, i.e., the probability of identifying effect heterogeneity if none exists and retains the underlying subgroups. This feature makes our technique particularly appealing in the context of clinical trials, where there may be significant costs associated with erroneously declaring that effects differ across population subgroups. TEHTrees are able to identify heterogeneous subgroups, characterize the relevant subgroups and estimate the associated treatment effects. We demonstrate the efficacy of the proposed method using a comprehensive simulation study and illustrate our method using a nutrition trial dataset to evaluate effect heterogeneity within a patient population. 
}\\
{Treatment effect heterogeneity; causal effects; conditional inference trees; matching.}
\end{abstract}

\section{Introduction \label{section intro}}

Under mild assumptions, randomized experiments estimate the average causal effect (ACE) of an intervention, also referred to as the average treatment effect (ATE). However, individuals may vary in their response to intervention so that the ATE is a poor representation of some individuals' expected benefit (or harm) from the intervention, a phenomenon often referred to as treatment effect heterogeneity \citep{Longford1999}.
%, Gabler2009, kravitz2004evidence}. 
%Characterizing treatment effect heterogeneity can reveal both ``weak responder'' and ``strong responder'' subpopulations, leading to more precise tailoring of intervention strategies.
For a given intervention, two heterogeneity-related questions arise: 1) Is the effect of the intervention heterogeneous? 2) If the intervention effect is heterogeneous, how does it vary across individuals? Many heterogeneity-focused statistical methods address both questions using a single model or procedure. Traditional approaches to characterizing treatment effect heterogeneity have primarily centered around regression modeling with interaction terms between the treatment and covariates. In such models, interaction terms can be used to assess whether treatment effect heterogeneity exists and also to characterize its magnitude.  Alternatively, formal nonparametric tests have been developed to test a null hypothesis of zero average treatment effect for any subpopulation defined by covariates and whether the average treatment effect is identical for all subpopulations \citep{Crump2008}. 

\subsection{Existing methods for the identification of treatment effect heterogeneity}

A rapidly growing toolkit of flexible machine learning techniques has produced several data-driven methods for characterizing treatment effect heterogeneity \citep{lipkovich2017, loh2019, Goldstein2019, dorie2019}. Flexible methods for heterogeneous treatment effect estimation include methods based on random forests \citep{athey2019, Wager2017, Su2018}, the LASSO \citep{Imai2013}, recursive partitioning \citep{su2009, Foster2011, athey2016}, boosting \citep{Powers2018}, Bayesian frameworks \citep{gk12, hill2011, hahn2018, Hahn2020}, combination frameworks \citep{kunzel2019, luedtke2016}, a generic machine learning approach \citep{chernozhukov2018} and a general class of two-stage algorithms that seek to address concerns about adapting machine learning for effect estimation \citep{nie2017}.

While many of the aforementioned techniques have shown impressive abilities to identify heterogeneous subgroups in situations where heterogeneity exists, they are often overly aggressive in identifying treatment effect heterogeneity when, in truth, there is none. Put more simply, most existing procedures for detecting treatment effect heterogeneity do not control Type I error. This lack of control of Type I error is particularly problematic in the context of randomized trials, where false declarations of treatment effect heterogeneity for a therapeutic agent could lead to wasteful follow-up studies and inappropriate off-label use in identified sub-populations. Several recently-proposed approaches have taken initial steps to address this issue: \citet{zhang2018} proposed two new splitting criteria within a CART-like framework to maintain a balance between minimizing error in estimating the treatment effect and maximizing heterogeneity; \citet{watson2020} compiled machine learning approaches to formally test for the presence of effect heterogeneity, evaluate the predictive benefits of the machine learning algorithm over a traditional statistical model and exert control over the type I error rate; \citet{Foster2011} recommended computing an associated measure of uncertainty for each subgroup to account for the consequences of false discovery and \citet{rigdon2018} introduced a matching plus classification and regression tree (mCART) that reduces the potential for falsely detecting treatment effect heterogeneity. Other approaches (e.g., \cite{Wager2017}) have also offered a framework to formally test for the overall effect heterogeneity. However, no flexible, unified framework has yet emerged that combines formal inferential testing for treatment effect heterogeneity with a focus on characterization of subgroups experiencing differential treatment effects.

\subsection{Our contribution}

In this paper, we propose a novel approach to testing for and characterizing effect heterogeneity of a randomized binary treatment on a continuous outcome, while explicitly controlling the Type I error rate. Our Treatment Effect Heterogeneity Tree (TEHTree) method involves building a conditional inference tree using pairs of individuals matched on the prognostic score \citep{h08}. After describing the TEHTree method and providing theoretical motivation for matching based on prognostic scores, we present the results of a substantial simulation study demonstrating TEHTree's Type I error control (power) in the absence (presence) of effect heterogeneity and compare its performance to other established techniques. This simulation study also evaluates subgroup identification and associated characterization for a unified framework (as in Causal Tree and TEHTree) against a two-stage method, which first generates individual treatment effects from an ensemble framework and then runs the effects through a conditional inference tree to determine subgroups. We also offer a comparison of the real-world performance of the two methods using data from a recent randomized trial in nutrition. %In the mCART framework, matched pairs are formed by sampling with replacement and within-pair differences in the outcome are correlated even when the original observations are independent. However, such matched pairs within the standard conditional inference tree framework inflate the Type I error.

\section{Method}
\subsection{Setup and Notation}

%\AVnote{The authors mention the SUTVA and note that these assumptions may not be reasonable for studies of infectious diseases in closed populations. While this statement is fairly intuitive, it might be good to include a bit more detail on why this is the case.}

Let $\mathbf{Y}=\{Y_{1},\dots ,Y_{n} \}$ be a continuous response vector for $k$ subjects randomized to treatment $Z=0$ and $n-k$ subjects randomized to treatment $Z=1$. The treatment assignments for all subjects are denoted by $\mathbf{Z}=\{Z_1,\dots,Z_n\}$. An accompanying $n \times p$-dimensional matrix $\mathbf{X}=\{\mathbf X_1,\dots,\mathbf X_p \}$ contains the $p$ covariates for each of the $n$ subjects with $\mathbf{X}_i=\{X_{i1}, \dots, X_{ip} \}$. 

In the counterfactual framework, each individual has a pair of counterfactual outcomes $(Y_{0i}, Y_{1i})$, where $Y_{1i}$ is the outcome of subject $i$ if assigned to treatment $Z=1$ and $Y_{0i}$ the outcome if assigned to $Z=0$. Hence, every individual has a (counterfactual, causal) treatment effect $Y_{1i} - Y_{0i}$, and the Average Treatment Effect (ATE) is defined as the mean of these within-individual differences, $ATE = E[Y(1) - Y(0)]$ (we have switched to the parenthetical counterfactual notation $Y(Z)$ to denote the observation $Y_{Zi}$ for arbitrary $i$). One of the benefits of randomization is that, under often plausible assumptions, the difference in means of randomized groups $E(Y|Z=1) - E(Y|Z=0)$ estimates the ATE. The Stable Unit Treatment Value Assumption (SUTVA) \citep{Rubin1980} bundles together two assumptions: 1) the treatment assigned to an individual affects only the outcome for that individual, and 2) there is only one ``version'' of treatment. In most randomized studies, SUTVA is plausible, a notable exception being studies of infectious diseases in closed populations. The other key assumption is ignorability, i.e., that treatment assignment $Z$ is independent of the counterfactual pair $(Y(0), Y(1))$. While this ``no unmeasured confounding'' assumption is non-trivial in observational studies, it is satisfied by design in a randomized trial.

%A central aspect of infectious diseases is that whether an individual is infected or not is dependant on who else is infected and the infection outcome in one patient is dependant on the vaccination status of the other patient, this immediately violates the first assumption (of no interference). 

%The other key assumption is ignorability, i.e., that treatment assignment $Z$ is independent of the counterfactual pair $(Y(0), Y(1))$. While this ``no unmeasured confounding'' assumption is non-trivial in observational studies, it is satisfied by design in a randomized trial.

The counterfactual framework allows every individual $i$ to experience a different effect of treatment, but because study participants are typically assigned to either $Z=1$ or $Z=0$, these individual-level effects $Y_{1i} - Y_{0i}$ are unobserved. Instead, we can characterize treatment effect heterogeneity in the counterfactual framework by estimating the Conditional Average Treatment Effect \citep{abadie2002},
\[
CATE(\mathbf{x}) = E(Y(1) - Y(0) | \mathbf{X} = \mathbf{x}).
\]
Because $Z$ is randomized, ignorability holds within any subset defined by $\mathbf{X} = \mathbf{x}$, and hence $CATE(\mathbf{x})$ can be estimated from $E(\mathbf{Y}|Z=1,\mathbf{X} = \mathbf{x}) - E(\mathbf{Y}|Z=0, \mathbf{X} = \mathbf{x}) \equiv \mu_1(\mathbf{x}) - \mu_0(\mathbf{x})$.  Therefore, the key challenge to characterizing treatment effect heterogeneity in randomized studies is to identify distinct subgroups defined by $\mathbf{X}$ with different CATEs. In the absence of treatment effect heterogeneity, the null hypothesis  $H_0: CATE(\mathbf{x}) = ATE  \ \forall \mathbf{x}$ holds. Most methods that seek to characterize how CATEs employ flexible semi- and non-parametric techniques in an attempt to identify regions of heterogeneity, but do not control the Type I error probability. In contrast, our approach embeds a classical parametric regression framework within a flexible tree model, allowing for both explicit control of the Type I error rate and characterization of treatment effect heterogeneity when it exists. The following two sections introduce the matching and conditional inference tree techniques that form the basis of our method.  %treatment is randomized, simple differences of averages (both overall and within subgroups defined by $\mathbf{X}$) are unbiased for the ATE, a fact that is important for our technique. \JWnote{Need to think here about whether we want to mention observational studies at all. I'm not sure our technique really handles potenital confounders that well.} %The prediction rule will be denoted by $d(X_i)$, and let the risk be defined by $r^* (d)=E[L(Y,d(X_i ))]$ where $L$ is the loss function of interest. For a continuous response, it is common to take $L(Y,d(X_i ))=(Y-d(X_i ))^2$, and the tree-based prediction rule $d(X_i )$ is referred to as a regression tree. 

\subsection{Matching}

If we observed $Y_{1i}$, $Y_{0i}$, and $\mathbf{X}_i$ for all $i$ then standard approaches to characterizing variability in a continuous outcome with respect to covariates could be used to estimate CATEs; for example, we could fit a regression tree using the differences $Y_{1i} - Y_{0i}$ as outcomes and $\mathbf{X}_i$ as predictors. However, in most trials an individual's outcome is observed under only one treatment, and hence $Y_{1i} - Y_{0i}$ is unobserved. So, we propose to impute it by matching each individual $i$ assigned to $Z_i = 1$ with a ``similar'' individual $j$ having $Z_j = 0$ and using $(Y_i - Y_j)$ to approximate $Y_{1i} - Y_{0i}$. If $j$ is an ``exact'' match for $i$ in the sense that $\mathbf{X}_i = \mathbf{X}_j$, then we can use $E(Y_i - Y_j | \mathbf{X}_i = \mathbf{x})$ to estimate $CATE(\mathbf{x})$. When the number of covariates $p$ is even moderately large and/or elements of $\mathbf{X}$ are continuous, it will typically be impossible to find exact matches for most individuals. One way of overcoming this problem is by deriving a single measure that characterizes the ``distance'' between individuals. If two individuals $i$ and $j$ with $Z_i = 1$ and $Z_j = 0$ have distance $d_{ij} = d$ between them,
\begin{equation}
    E(Y_i - Y_j | \mathbf{X}_i = \mathbf{x}, \mathbf{X}_j = \mathbf{x}', d_{ij} = d, Z_i = 1, Z_j = 0) =  CATE(\mathbf{x}) + \Delta^0_{ij}(\mathbf{x}, \mathbf{x}', d) \label{eq:cate-delta}
\end{equation}
\noindent (see Appendix A for the short proof). Hence, pairs matched according to $d_{ij}$ can be used to estimate CATEs provided $\Delta^0$ is small. Note that, in general, $\Delta^0$ may be non-zero even if $d = 0$; it is the price paid for reducing the multidimensional vectors $\mathbf{x}$ and $\mathbf{x}'$ to the scalar distance $d$. 

A number of distance measures for matching have been proposed, some of which we review briefly here. Broadly speaking, these measures can be broken down into three categories according to how they define similarity: based on the distance between covariate vectors (e.g., Mahalanobis distance), based on the probability of being treated (propensity score), and based on the predicted value of the outcome (prognostic score). %Covariate-based distances such as the Mahalanobis distance are vulnerable to the dimensionality reduction penalty alluded to above; when the number of covariates is large, $\Delta^0(\mathbf{x}, \mathbf{x}', d = \epsilon)$ may be relatively large even if the scalar distance $\epsilon$ is small (or zero) \citep{Gu1993}. 
%The propensity score , defined as the probability of receiving the treatment given the observed covariates, is often used for matching. 
The propensity score \citep{rr83, Stuart2013} is unhelpful, when treatment is randomized, since by design the covariates are independent of treatment assignment and as a result propensity score matching does not make $\Delta^0$ small. Another way of summarizing the similarity between individuals is via the prognostic score \citep{h08}. %If $\phi(\mathbf{X})$ is sufficient for $Y_0$, in the sense that $Y_0$ is independent of $\mathbf{X}$ given $\phi(\mathbf{X})$, we call $\phi(\mathbf{X})$ a prognostic score. 
Individuals with similar prognostic scores have similar predicted values of the outcome under treatment $Z=0$ (typically a control condition). %If $\mathbf{Y}|\mathbf{X}$ follows a homoscedastic linear model, then $E(Y_0|\mathbf{X})$ is a prognostic score, which can be estimated by fitting a regression of $\mathbf{Y}$ on $\mathbf{X}$ among those with $Z=0$, then using that model to obtain predictions of the outcome under the control condition for all individuals \citep{Stuart2013}. 
Matching on prognostic scores is appealing in our context where the goal is approximate individual causal treatment effects $Y_{1i} - Y_{0i}$. %As noted above, when $Y_0|\mathbf{X}$ follows a homoscedastic linear model, $E(Y_0|\mathbf{X}) \equiv \phi(\mathbf{X})$ is a prognostic score. 
If two individuals $i$ and $j$ with $Z_i = 1$ and $Z_j = 0$ have the same prognostic score $\phi(\mathbf{X}_i) = \phi(\mathbf{X}_j) = \phi$, 
\begin{equation}
E(Y_i - Y_j \mid \mathbf{X}_i = \mathbf{x}, \mathbf{X}_j = \mathbf{x}', \phi, Z_i = 1, Z_j = 0) = CATE(\mathbf{x}) \label{eq:cate}
\end{equation}
(see proof in Appendix A). This result immediately implies that if $d_{ij} = |\phi(\mathbf{X}_i) - \phi(\mathbf{X}_j) | = 0$, then $\Delta^0 = 0$ in Equation \ref{eq:cate-delta}. In other words, matching on the prognostic score yields pairs that can be used to estimate conditional average treatment effects. Note that this result holds if $\mathbf{X}$ is replaced by any measurable function $m(\mathbf{X})$, so that if $m$ captures the way in which $\mathbf{X}$ modifies the effect of treatment, then pairs matched on $\phi(\mathbf{X})$ retain information about effect modification.

\subsection{Conditional Inference Trees}

With matched pairs in hand that can be used to estimate CATE$(\mathbf{x})$, the next step is to characterize how the CATE varies with $\mathbf{x}$. Our approach uses  conditional inference trees, a variant of decision trees which we briefly introduce here.%As noted above, we seek a method that will describe effect heterogeneity where it exists, but control the probability of identifying variation in the CATE when in truth it does not depend on $x$. We therefore adopt the method of conditional inference trees, which we briefly introduce here.

%Decision trees \citep{ms63} are non-parametric models based on recursively partitioning a sample into distinct subgroups that share similar outcome values. Sample partitions (or ``splits'') are defined by thresholding covariate values. 
The most popular and commonly used technique for building decision trees, the Classification and Regression Tree (CART) technique, was introduced by \cite{Breiman2017} (originally 1984). Because of its interpretability and flexibility, CART has also been incorporated into several methods for assessing treatment effect heterogeneity. For example, the Causal Tree \citep{athey2016} optimizes for heterogeneity in treatment effects and uses a modified mean-squared criterion expression within a CART framework for both splitting and cross-validation.
One of CART's drawbacks is that, because it considers many possible thresholds on all possible variables when searching for an optimal split, it has a tendency to overfit the data on hand and produce overly complex models. This overfitting tendency can be controlled somewhat by ``pruning'' trees based on a complexity parameter. However, as we show in our simulation study, even pruned CARTs do not control the Type I error for effect heterogeneity. Several related methods have demonstrated excellent performance in detecting the presence or absence of underlying treatment effect heterogeneity and estimating individual-level CATEs. For example, the Causal Forest algorithm \citep{Wager2017}, which was developed as an ensemble over Causal Trees, offers an associated formal test for the presence of effect heterogeneity and can estimate $CATE(\mathbf{x})$ for a wide variety of functional relationships between $\mathbf{x}$ and the treatment effect. However, this and similar ensemble methods do not generally yield an interpretable set of subgroups characterizing effect heterogeneity; we highlight these differences in characterization in our simulation study in Section \ref{sect:simstudy}.

One alternative to a CART framework is the Conditional Inference Tree (CTree), proposed by \cite{hhz06}. The main difference between CTrees and CARTs is in the splitting process: in CTrees, the processes for choosing a variable to split on or to stop splitting (the ``variable selection'' step) and choosing an optimal splitting threshold for the selected variable (the ``splitting'' step) occur sequentially, while in CART they happen simultaneously. In the variable selection step of CTree, the decision of whether not to continue splitting is based on a test of the global null hypothesis $H_0: E(\mathbf{Y}|\mathbf{X})=E(\mathbf{Y})$, which is tested by considering all marginal null hypotheses $H_0^m: E(\mathbf{Y}|\mathbf{X}_m) = E(\mathbf{Y})$ for $m=1,\dots,p$. In a simple case, each $H_0^m$ can be assessed by calculating the p-value for the slope term from a univariate regression model of $\mathbf{Y}$ on $\mathbf{X}_m$. More generally, this step can accommodate a wide variety of models and test statistics; even if a statistic's sampling distribution is unknown, permutation tests can be used to calculate p-values for each partial null hypothesis. 

Since the global null $H_0=\bigcap_{m=1}^p H_0^m$ is rejected if the minimum p-value for all of the partial null hypotheses is less than a pre-specified level of significance, control of Type I error can be achieved by setting this level using an appropriate multiplicity adjustment to account for the testing of the $p$ partial null hypotheses (see Section \ref{sect:testing}). If the minimum p-value exceeds the threshold, the tree does not split the given subset further. Otherwise, the partial null hypothesis $H_0^s$ that results in the smallest p-value will indicate the covariate $\mathbf{X}_s$ that is most strongly associated with the outcome $\mathbf{Y}$ and the algorithm proceeds to the next step to determine how to optimally threshold $\mathbf{X}_s$. In our method, we calculate p-values associated with the (fixed) slope term from univariate linear mixed models. %As we discuss below, the choice of model to calculate p-values in the variable selection step determines which types of associations will generate splits and which will not, i.e., what forms of effect heterogeneity can be detected. 
Once covariate $\mathbf{X}_s$ has been selected for splitting, the second step of the CTree algorithm is to find the threshold $c$ that maximizes the discrepancy $|E(\mathbf{Y}|\mathbf{X}_s\leq c) - E(\mathbf{Y}|\mathbf{X}_s > c)|$. %Typically, the process is performed over a finite set of candidate splits $\mathbf{c} = \{c_1, \dots, c_k\}$ which may be restricted to ensure properties such as a minimum number of observations in the partitions defined by $(\mathbf{X}_s \leq c)$ and $(\mathbf{X}_s > c)$. 
This two-phase splitting procedure is repeated on the resulting partitions until no more subsets are eligible for splitting. 

\subsection{Treatment Effect Heterogeneity Trees (TEHTrees)}
\label{sect:algorithm}

We propose a two-stage approach to assessing treatment effect heterogeneity in randomized studies. In the first stage, prognostic scores are calculated and every treated subject is matched to a control subject (with replacement) based on the prognostic score. %In the second stage, a training sample composed of the outcome and covariate values for the treated subject in the matched pair is used as inputs to a conditional inference tree. A separate sample is then used to estimate the differential treatment effects across the subgroups determined by the conditional inference tree structure. 
In the second stage, within-pair differences in the outcome along with the covariate values of the treated member of each pair are used as inputs to a conditional inference tree. The nodes of the fitted conditional inference tree identify subgroups across which the causal effect of treatment varies. 
The full algorithm is as follows; in the sections that follow, we provide details about its key steps.

\subsubsection{TEHTree algorithm}

\begin{enumerate}
\item{Separate the dataset into a training and holdout set for the purpose of constructing a tree (steps 2-5 below) and for estimating treatment effects (step 6).}
\item Fit a model to calculate prognostic scores $\phi$ using individuals in the training data with treatment status $Z=0$, and obtain estimated prognostic scores $\hat \phi$ for each individual in the sample. Model details for prognostic score estimation are provided in Section \ref{sect:prognostic}.
\item Form a set of matched pairs $\{(i_1, j_1), (i_2, j_2), \dots, (i_k, j_k)\}$ from the training data by matching each treated ($Z = 1$) subject with one control ($Z = 0$) subject, with replacement, based on $\hat \phi$. Ties are broken randomly.
\item For each pair $(i_l,j_l)$, calculate the within-pair difference in the outcome, $\delta_l = Y_{i_l} - Y_{j_l}$. Each pair can now be viewed as a single ``pseudo-individual'' represented by the scalar continuous outcome $\delta_l$ and the covariate vector $\mathbf{X}_l \equiv \mathbf{X}_i$.
\item Use the pseudo-individual data $(\delta_l, \mathbf{X}_l)$ created in the previous step and the desired Type I error rate as inputs to create a Treatment Effect Heterogeneity Tree (TEHTree), as described in Section \ref{sect:testing}. 
\item Estimate the treatment effect within each terminal node of the fitted TEHTree as described in Section \ref{sect:estimation}.
\end{enumerate}

The preceding algorithm assumes that a sufficient amount of data is available to create a holdout test set of sufficient size to accurately estimate treatment effects within subgroups defined by each terminal node. If the sample size is limited, steps 2-6 can be carried out on the entire dataset with a single-sample estimation approach for step 6 as described in Section \ref{sect:estimation}. 

\subsubsection{Estimating prognostic scores}
\label{sect:prognostic}

To provide robustness against misspecification of the prognostic score model, we apply the Super Learner \citep{Van07} to estimate the prognostic score $\phi(\mathbf{X}) = E(\mathbf{Y}|Z=0,\mathbf{X})$ using data from the untreated ($Z=0$) group. The base learners in our application consist of the sample mean, a linear model (with and without interaction terms), a generalized additive model, a random forest, stepwise regression (with and without interaction terms), and ``polymars'' (multivariate adaptive polynomial spline regression) as base learners. %The Super Learner estimates the performance of these base learners using cross-validation and develops a weighted average over a combination of these methods. 

\subsubsection{Testing partial null hypotheses}
\label{sect:testing}

Because the outcome values $\delta_{i,j}$ are derived from pairs formed by matching with replacement, inputs to the conditional inference tree are correlated and hence a standard univariate linear regression-based approach to evaluating the partial null hypotheses $\{ H^0_m: E(\mathbf{Y}|\mathbf{X}_m) = E(\mathbf{Y}) \}$ will produce invalid p-values. Instead, we test partial null hypotheses by fitting univariate linear mixed models of the form:
\begin{equation}
E(\delta_{i,j} | X_{im}, b_j) = \beta_0 + \beta_1 X_{im} + b_j \label{eq:mixedmodel}
\end{equation}
where $b_j \sim N(0, \tau^2)$ is a random intercept corresponding to the control subject in each pair. Similar models are used to determine the optimal splitting for selected covariate $\mathbf{X}_s$, replacing $X_{im}$ in Equation \ref{eq:mixedmodel} by $\mathbf{1}[X_{is} \geq c]$.

To establish proof of concept for our method, we used the Bonferroni method to adjust the marginal hypothesis test p-values for multiple comparisons, which sets the significance threshold at $\frac{\alpha}{p}$ for desired Type I error rate $\alpha$. Other less conservative adjustment methods could also be applied; see the Discussion for more details. 

%\AVnote{In Section 2.4.3, it might be good to modify the notation slightly to make it more clear that only patients in the test data (who belong to a particular terminal node) are used to compute estimated treatment effects within that terminal node.}

\subsubsection{Treatment effect estimation}
\label{sect:estimation}

%A TEHTree assigns every observation to a terminal node in the tree; if the tree has more than one terminal node, we conclude there exists heterogeneity in treatment effects because the estimated treatment effect differs between these terminal nodes. 
We propose a double-sample (test/holdout set) approach to estimating heterogeneous treatment effects using TEHTree that parallels the one used in the Causal Tree method. After having generated the TEHTree based on the training set, we determine the TEHTree terminal node that each individual in the holdout set belongs to. Each terminal node $\mathcal{T}$ consists of the union of two subsets $\mathcal{T}_1 = \{i: Z_i = 1, i \in \mathcal{T}\}$ and $\mathcal{T}_0 = \{i: Z_i = 0, i \in \mathcal{T}\}$. We compute the treatment effect estimate as $\Delta(\mathcal{T}) = \frac{1}{|\mathcal{T}_1|} \sum_{\mathcal{T}_1} Y_i^{te} - \frac{1}{|\mathcal{T}_0|} \sum_{\mathcal{T}_0} Y_i^{te}$, where $Y_i^{te}$ refers to outcomes in the test/holdout set. Note that, for a 1:1 randomized treatment, large discrepancies between $|\mathcal{T}_1|$ and $|\mathcal{T}_0|$ are unlikely, and hence the precision of $\Delta(\mathcal{T})$ will be approximately proportional to $\frac{1}{\sqrt{|\mathcal{T}|}}$.

If limited sample size precludes carving out an independent holdout set from the original data, a straightforward single-sample estimation approach can be used. Let $\mathcal{T}$ denote the set of matched pairs from the data belonging to each terminal node.  Then, the single-sample treatment effect estimate is simply $\Delta(\mathcal{T}) = \frac{1}{|\mathcal{T}|} \sum_{l \in \mathcal{T}} \delta_l$. Estimation of the precision of the single-sample estimate of $\Delta(\mathcal{T})$ is complicated (relative to the double-sample approach) by the need to consider the correlation between matched pairs. In Section \ref{sect:illustration}, we apply this single-sample approach to the data illustration in the presence of a limited sample size.

%The Causal Tree uses MSE-based criterion expressions to build (both splitting and cross validation) the tree structure and estimate the conditional average treatment effect at each leaf node over covariate-defined subpopulations. 
%A key feature of the Causal Tree approach is that it partitions the data into two subsets, one of which is used for subgroup construction and the other of which is used to estimate treatment effects within these subgroups. 

\subsection{Implementation}

We implemented TEHTree in R \citep{t18} using a modified conditional inference tree framework and relevant functions in the \texttt{partykit} package \citep{hrn2015}. Matching was conducted using the \texttt{Matching} package \citep{s11}, and all linear mixed models were fit using the \texttt{nlme} package \citep{pbdst}. The \texttt{Super Learner} was used to estimate the prognostic score. Code for implementing TEHTree can be found at \texttt{https://github.com/AshwiniKV/TEHTree}. The Causal Tree and Causal Forest were implemented using the \texttt{causalTree} and \texttt{grf} packages \citep{tibshirani2018}.

\section{Simulation Study}
\label{sect:simstudy}

%\AVnote{Include more details about the default settings of Causal trees}

We conducted simulations to evaluate the TEHTree method and compare its performance to other approaches, including several variants of the Causal Tree technique. We evaluated the Type I error, power, and other statistical properties of the techniques for the tree types over different data generating scenarios that are documented in the Appendix (Tables \ref{tab:sim_parameters} and \ref{tab:sim_scenarios}). Factors that were varied over the scenarios include sample size ($N$ = 100, 200, 500, 1000 and 2000), number of covariates, type of covariates (binary and continuous), coefficients, and pairwise correlation among covariates. The treatment variable $Z$ was generated such that $N/2$ subjects received treatment ($Z=1$) and $N/2$ subjects received control ($Z=0$). All results described in this simulation study are based on 1,000 simulations per scenario. Continuous covariates were generated from multivariate normal distributions with mean zero, unit variance, and varying pairwise correlations. Binary covariates were generated as independent Bernoulli$(0.5)$. Continuous outcomes were generated as independent Normal with unit variance and means depending on the scenarios detailed in Appendix B.

%The settings for the tree function include a splitting rule that follows the honest splitting approach within a Causal Tree framework and the fit option within the cross validation option but does not utilise the honest risk evaluation approach within cross validation.  implemented using the package \texttt{causalTree} \citep{ctpkg2016}

\subsection{Type I Error}

\begin{figure}[ht]
    \centering
    \includegraphics[scale = 0.52]{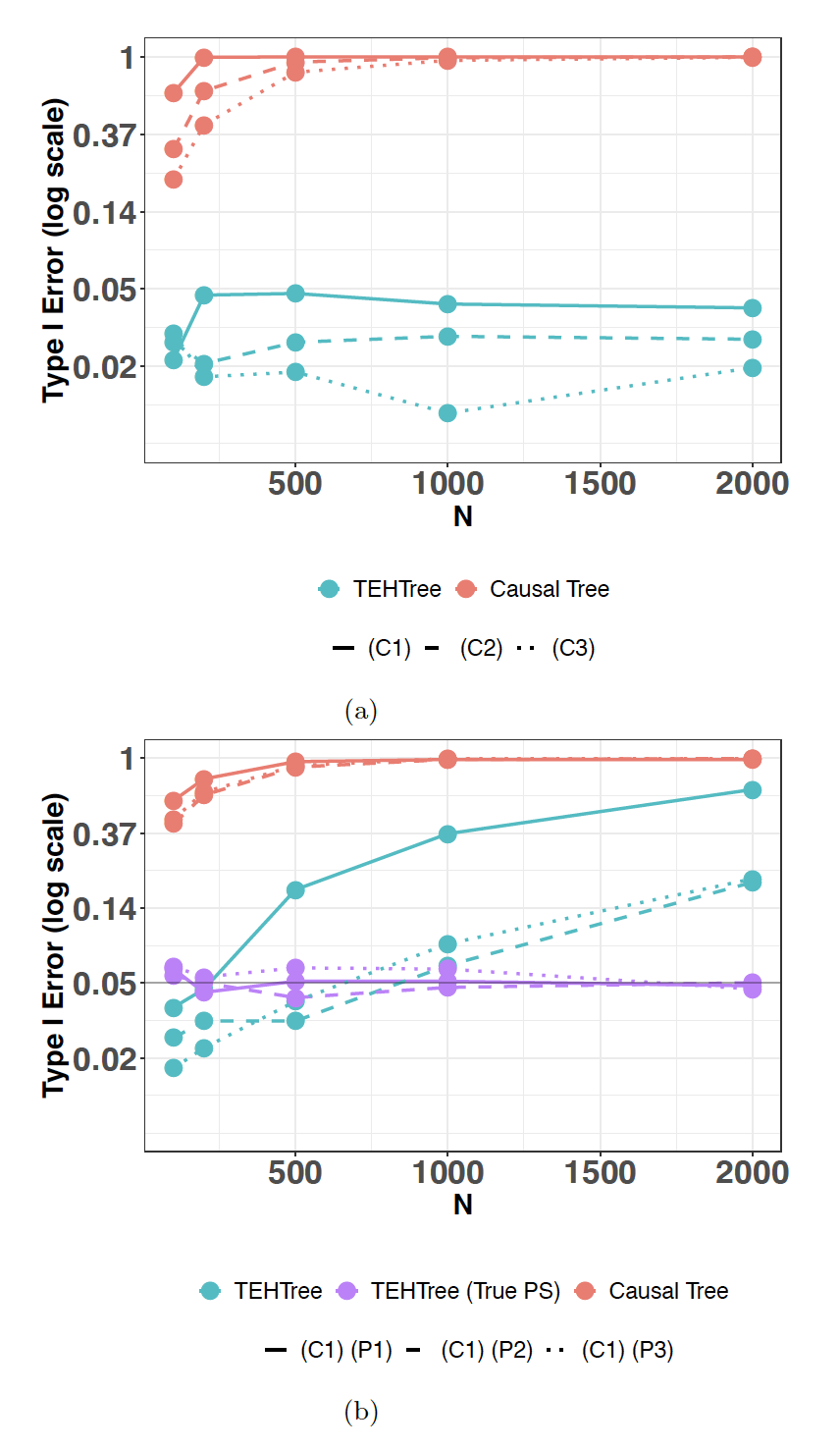}
    \caption{Results of simulations conducted over different scenarios to evaluate the Type I Error rate for TEHTrees and Causal Trees. The plots use data generated from Models M1 and M2, as specified in the Appendix (Table \ref{tab:sim_parameters}). The different scenarios associated with each model and their parameters are also further summarised in the Appendix (Table \ref{tab:sim_scenarios}). %by models including main effects for treatment and covariates, for different types and number of covariates (5 Binary, 5 Continuous and 10 Continuous), vectors of coefficients and across sample sizes. 
    The error rates are displayed at sample sizes $N$ = 100, 200, 500, 1000 and 2000.}
    \label{fig:plots1}
\end{figure}

For a case when there is no treatment effect heterogeneity, we will say that a tree-based method for detecting heterogeneity has committed a Type I error when the tree generates more than one terminal node, incorrectly implying that treatment heterogeneity exists. We generated data under two sets of scenarios where there was no treatment effect heterogeneity. In the first set of scenarios, outcomes were generated from a linear model with main effects for treatment and covariates (Model (M1) in the Table \ref{tab:sim_parameters}). In these scenarios, a simple linear model including treatment and covariates correctly specifies the prognostic score; in other words, the SuperLearner ensemble used to estimate the prognostic score contains the correct model. Figure \ref{fig:plots1}(a) displays the Type I error (in logarithmic scale) of TEHTrees and Causal Trees for data simulated under these scenarios. In all cases, the Type I error of TEHTrees is less than the desired 0.05 level, while the Type I error of Causal Trees is greater than 0.05 in every scenario, usually substantially so. As the sample size increases, the Type I error of Causal Trees increases and is approximately 1 at $N = 2000$ for all three scenarios; the Type I error of TEHTrees stays roughly constant. In the second set of scenarios, outcomes were generated from a linear model with main effects of treatment and covariates, along with additional effects for thresholded versions of continuous covariates (Model (M2) in the Appendix, Table \ref{tab:sim_parameters}). In these scenarios, the SuperLearner ensemble does not contain the true model. Figure \ref{fig:plots1}(b) displays the Type I error (in logarithmic scale) for these scenarios. The Type I error rate of TEHTrees tends to increase with sample size and is no longer below the desired 0.05 in every scenario. However, the Type I error rate using TEHTrees is still much lower than the Type I error rate using Causal Trees. We note that this is a particularly challenging scenario for an approach based on prognostic score matching; even modest misspecification of the prognostic score could markedly increase the proportion of matched pairs where one individual has $X > 0$ and the other has $X \leq 0$, leading to the (erroneous) conclusion that treatment effects are heterogeneous in $X$. When TEHTree is used with with a correctly-specified prognostic score model (purple points and lines), the Type I error rate is once again controlled.   

\subsection{Power} 

 We characterized the performance of TEHTrees and Causal Trees under a number of different data generating scenarios where treatment effect heterogeneity is driven by binary (dichotomized) covariates only, and a mixture of binary and continuous covariates. We defined the power for detecting treatment effect heterogeneity as the probability that a tree produced a split on a variable having a non-zero interaction with treatment (i.e., one that is responsible for producing treatment effect heterogeneity). Note that while this definition of power does not credit trees with splitting on variables unrelated to heterogeneity, in cases where the treatment effect interacts with several different covariates, a tree is credited with rejecting the null hypothesis if it splits on any of these covariates. 
 
\begin{figure}[ht]
    \centering
    \includegraphics[scale = 0.489]{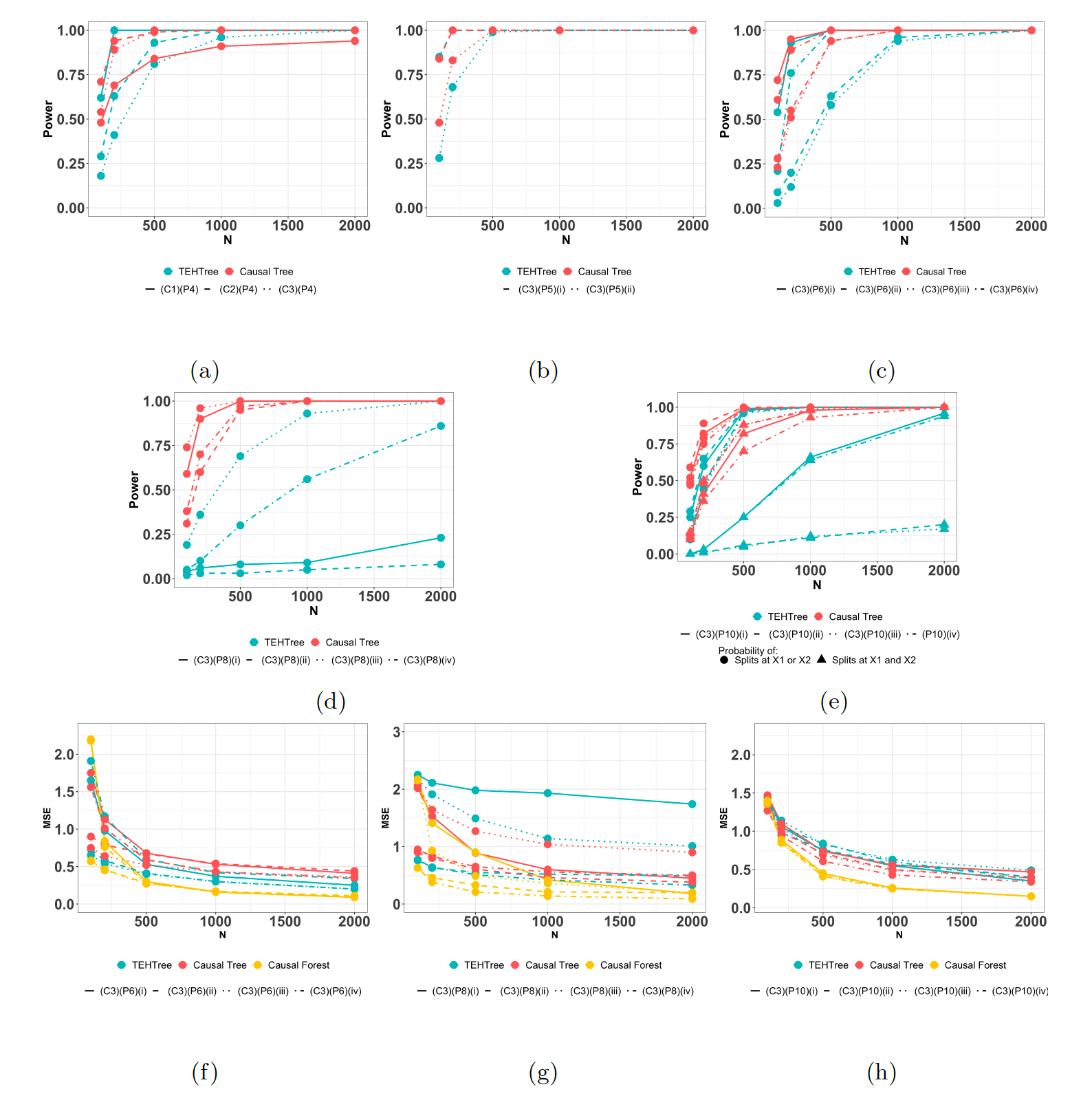}
    \caption{Figures \ref{fig:powersone}(a) - \ref{fig:powersone}(e) present values of power for TEHTrees and Causal Trees. The plots use data generated from Models M3, M4, M5, M7 and M9. Figures \ref{fig:powersone}(f) - \ref{fig:powersone}(h) presents the mean squared error (MSE) of the estimated average treatment effect. The plots use data generated from Models M5, M7 and M9. In both cases, data are generated according to different scenarios with sample sizes N = 100, 200, 500, 1000 and 2000. See Appendix for details of simulation settings and additional results.}
    \label{fig:powersone}
\end{figure}

 Figures \ref{fig:powersone}(a) through \ref{fig:powersone}(d) summarize the results of scenarios where heterogeneity is determined by a single covariate ($X_1$) (Models M3, M4, M5 and M7 in Appendix Table \ref{tab:sim_parameters}), and hence the power is the probability of splitting on $X_1$. TEHTrees and Causal Trees have similar power when heterogeneity is determined by $I(X_1 > 0)$ (Figure \ref{fig:powersone}(a)) and by continuous $X_1$ (Figure \ref{fig:powersone}(b)); the power of TEHTrees is modestly lower for small sample sizes when both the indicator and continuous value contribute to heterogeneity (Figure \ref{fig:powersone}(c)). TEHTrees and Causal Trees have similar power at higher sample sizes when heterogeneity is induced by $\sin(\eta X_1)$ (Figure \ref{fig:powersone}(d), at $\eta = 1.5$).
 
Figure \ref{fig:msepoweradd}(a) in the Appendix is notable as it shows that TEHTree has very low power when effect heterogeneity is driven by $I(-0.5 < X_1 < 0.5)$. This result is due to the fact that, in our implementation, the null hypothesis $H_0: E(\delta_{ij}|X) = E(\delta_{ij})$ is tested via the coefficient of the main effect of $X$ in a (mixed) linear model. When heterogeneity is due to the above indicator, this coefficient will often be estimated as being close to zero and hence the null hypothesis is unlikely to be rejected. An implementation which used a more robust test for variation of the mean with $X$ (e.g., by using a joint test for higher-order polynomial terms) would perform better in this case at the cost of greater computational complexity. As with the utilization of the appropriate functional form and its positive impact on performance, TEHTree is likely to have lower power if important predictor variables are simply not measured. 

Figure \ref{fig:powersone}(e) also displays the power of TEHTrees and Causal Trees when heterogeneity in the treatment effect is due to two variables, $X_1$ and $X_2$. For this Model M9, we considered two definitions of power: splitting on \emph{either} $X_1$ or $X_2$, and splitting on \emph{both} $X_1$ and $X_2$. In general, both types of power are higher for Causal Trees than TEHTrees. 

Results for other relevant Models M8 and M10 are available in the Appendix (Figure \ref{fig:msepoweradd}(b)-(c)). While the preceding results show that the Causal Tree approach has higher power than our proposed TEHTree method under many data generating scenarios, this comparison (like most power comparisons) is somewhat misleading since TEHTree controls the Type I error rate while Causal Tree does not. Figure \ref{fig:powercomp} (in the Appendix) shows how the power of Causal Tree and TEHTree vary in Models M4, M6 and M8 as the parameters that determine heterogeneity range from 0 (no heterogeneity) to larger values (substantial heterogeneity).

\subsection{Treatment Effect Estimation}
Figures \ref{fig:powersone}(f)-(h) present the mean squared error (MSE) of the treatment effect estimates across data generating scenarios and for different sample sizes. The scenarios are defined using Models M5, M7 and M9 and the details are fully described in the Appendix (Tables \ref{tab:sim_parameters} and \ref{tab:sim_scenarios}). Across most scenarios, Causal Trees and TEHTrees show similar MSEs and Causal Forests generally show lower MSEs at larger sample sizes. Characteristics of TEHTrees and Causal Trees using data generated by Model M3 and Model M7 are displayed in the Appendix (Figure \ref{fig:char}); the figure describes both the proportion of split points within a given distance of the true split point and the number of terminal nodes. 

\subsection{Subgroup identification}

\begin{figure}[htb!]
    \centering
    \includegraphics[scale = 0.5]{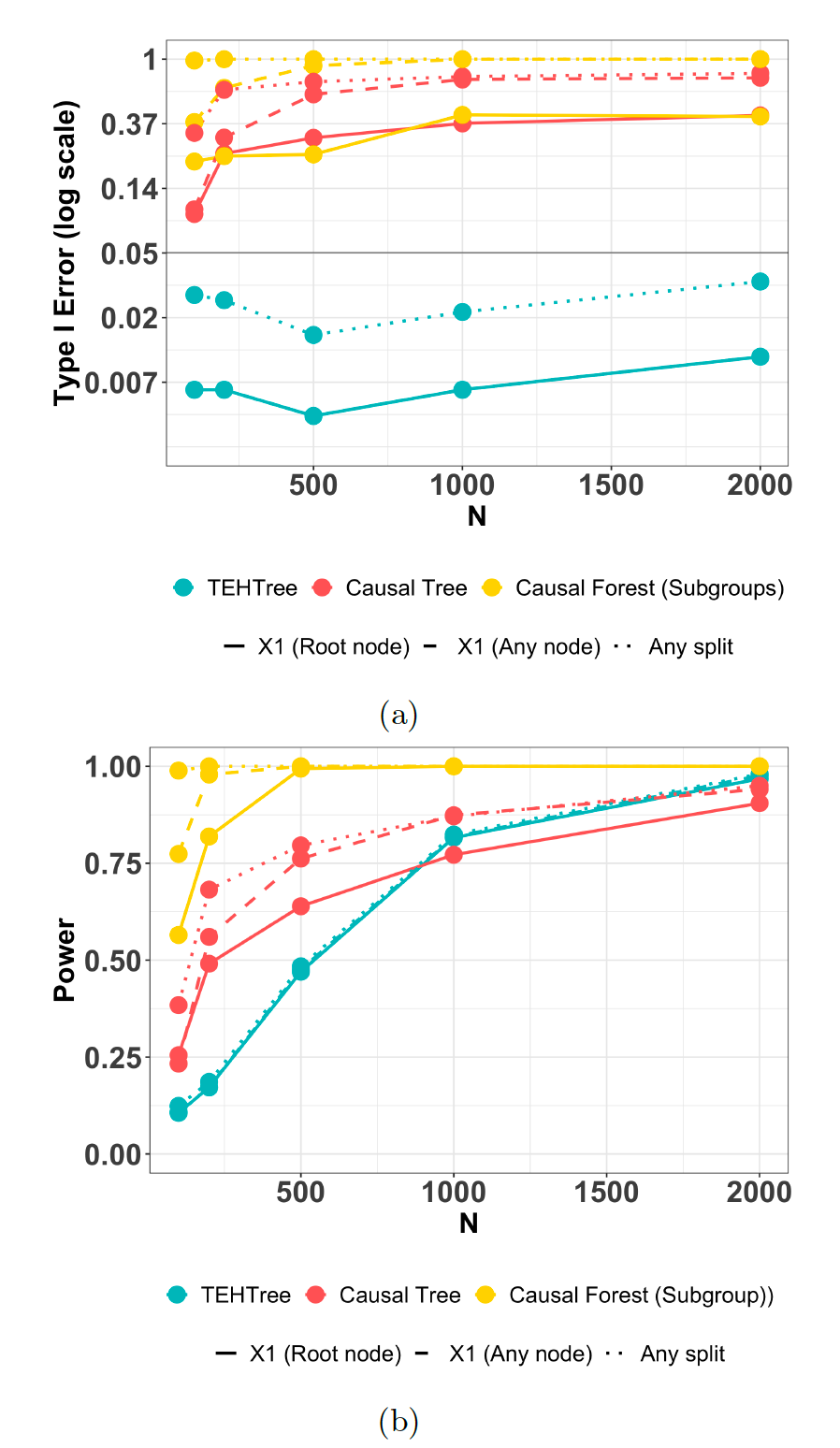}
    \caption{Type I error and Power of different tree types when data are generated over sample sizes N = 100, 200, 500, 1000 and 2000. The data to demonstrate the error rates and power are generated using Models M1 and M3 respectively. For a given simulation, these measures are defined over the probability of the tree splitting at X1 at the root node, the tree splitting at X1 at any node and the probability of a tree splitting at all.}
    \label{fig:subgroup}
\end{figure}

As demonstrated above, single-method techniques which do not control Type I error tend to "oversplit" and identify many subgroups when heterogeneity is truly defined by only a few. Figure \ref{fig:subgroup}(a) describes these differences in comparison to the Causal Tree and TEHTree algorithms using data generated from Model M1. Across sample sizes, TEHTree retains the Type I error rate control while Causal Forest generated effects leads to trees that continue to split in the absence of effect heterogeneity. Other simulation results for data generated from Models M4, M6, M7 and M8 can be found in the Appendix.

\begin{figure}[ht]
    \centering
    \includegraphics[scale = 0.6]{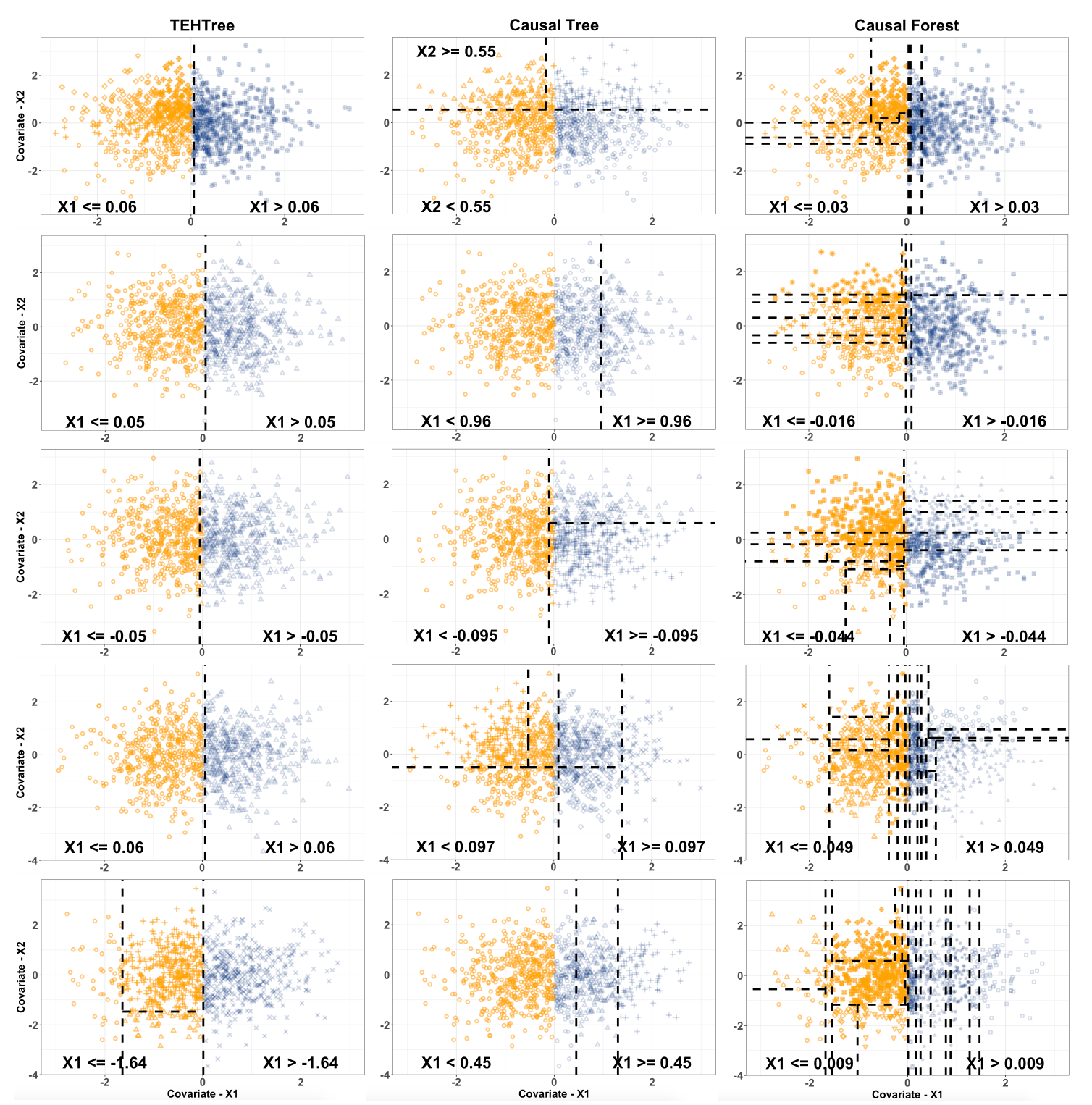}
    \caption{Series of plots showing partitions determined by different tree types (for a continuous outcome, continuous covariates and 1000 corresponding simulated observations using Model M3). The first split at each root node and the criterion is labelled within each plot. Navy blue points within each plot correspond to values where $X_1 \geq 0$ and orange points correspond to values where $X_1 < 0$. In the first row, the first plot displays partitions for a TEHTree (Root node splits at $X_1 \leq 0.06$), the second plot for a Causal Tree (Root node splits at $X_2 \geq 0.55$) and the third plot displays partitions extracted using a Conditional Inference Tree over treatment effects estimated from a Causal Forest (Root node splits at $X_1 \leq 0.03$). This panel displays partitions determined by three different tree types for five simulated datasets (as corresponding to the five rows and three columns). }
    \label{fig:subsplits}
\end{figure}

Figure \ref{fig:subsplits} illustrates this phenomenon by showing the sample partitions defined by the terminal nodes of three techniques: TEHTree, Causal Tree, and a decision tree with treatment effects from Causal Forest used as inputs.  %This Figure visualizes the differences in the number of splits between the tree types (i.e., Causal Tree, TEHTree and a conditional inference tree over Causal Forest's estimated treatment effects); 
Each column in the panel corresponds to a tree type and each row is a separate realization of a simulated dataset. In these data, the only driver of effect heterogeneity is $I(X_1 \geq 0)$, and the methods are given data on two covariates, $X_1$ and $X_2$. TEHTree consistently identifies the correct subgroups by splitting in the neighborhood of $X_1 = 0$; in four out of five simulated datasets, this is the only split TEHTree produces. The Causal Tree also regularly identifies a split near $X_1 = 0$, but also frequently identifies additional, non-heterogeneity-inducing splits on both $X_1$ and $X_2$. Treatment effects estimated by Causal Forest are much more heterogeneous, decision trees based on these effects produce multiple splits on both $X_1$ and $X_2$.

\section{Illustration} \label{sect:illustration}

We illustrate the application of TEHTree, Causal Tree, and Causal Forest to data from the Box Lunch Study (BLS) \citep{french2014a}, a randomized controlled trial to evaluate the effect of receiving daily boxed lunches of three different portion sizes (400 kcal, 800 kcal, and 1600 kcal) on daily energy intake and body weight of working adults. The BLS study enrolled 233 subjects including a group that did not receive any boxed lunches; we analyzed a complete-case version of the data where 156 subjects were assigned to receive one of the three fixed portion size lunches. More specifically, we consider the problem of characterizing heterogeneity across the effect of ``treatment'' (defined as the 800 and 1600 kcal boxed lunches, $n = 107$) versus ``control'' (the 400 kcal boxed lunch, $n = 49$) on daily caloric intake six months after randomization. The average treatment effect (ATE) is a difference of 193 kcal/day. For this analysis, we consider how the ATE varies with four baseline covariates: BMI, age, a measure of hunger, and EDEQ-14.0, a measure of loss of control over eating in the past 28 days.

%The set of baseline covariates in the BLS dataset that were available and may be related to effect heterogeneity include demographics (e.g., age, BMI, sex, education and job type), psycho-social measures (e.g., frequency of self-weighing, degree of satisfaction with current weight), responses to the Three Factor Eating Questionnaire (TFEQ) (e.g., measures of hunger, disinhibition and restraint), responses to the Eating Disorder Examination Questionnaire (EDEQ) (e.g., EDEQ-14.0 assesses the occurance of loss of control over eating in the past 28 days) and novel laboratory-based psycho-social measures such as relative reinforcement of food (rrvf), liking and wanting. 

We applied Causal Forest \citep{Wager2017} to estimate individual treatment effects and empirically evaluate the presence of effect heterogeneity. The Causal Forest was fit to a training dataset of 117 subjects including the 4 baseline patient characteristics and the relevant causal treatment effects were estimated on 39 test set subjects (displayed in Figure \ref{fig:plothte} in the Appendix). The results suggest that the subjects experience a relatively homogeneous response to treatment with respect to these covariates. The method does not explicitly identify covariates which induce heterogeneity. %Together, these plots help visualize the nature of effect heterogeneity within the trial dataset and the variance for these estimated treatment effects. More specifically, Figure \ref{fig:fullhte} indicates that the subjects experience a relatively homogeneous response to treatment and Figure \ref{fig:subhte} suggests the presence of treatment effect heterogeneity, albeit with greater uncertainity. 

%Figure \ref{fig:plothte} 

\begin{figure}[ht]
    \centering
    \includegraphics[scale = 0.475]{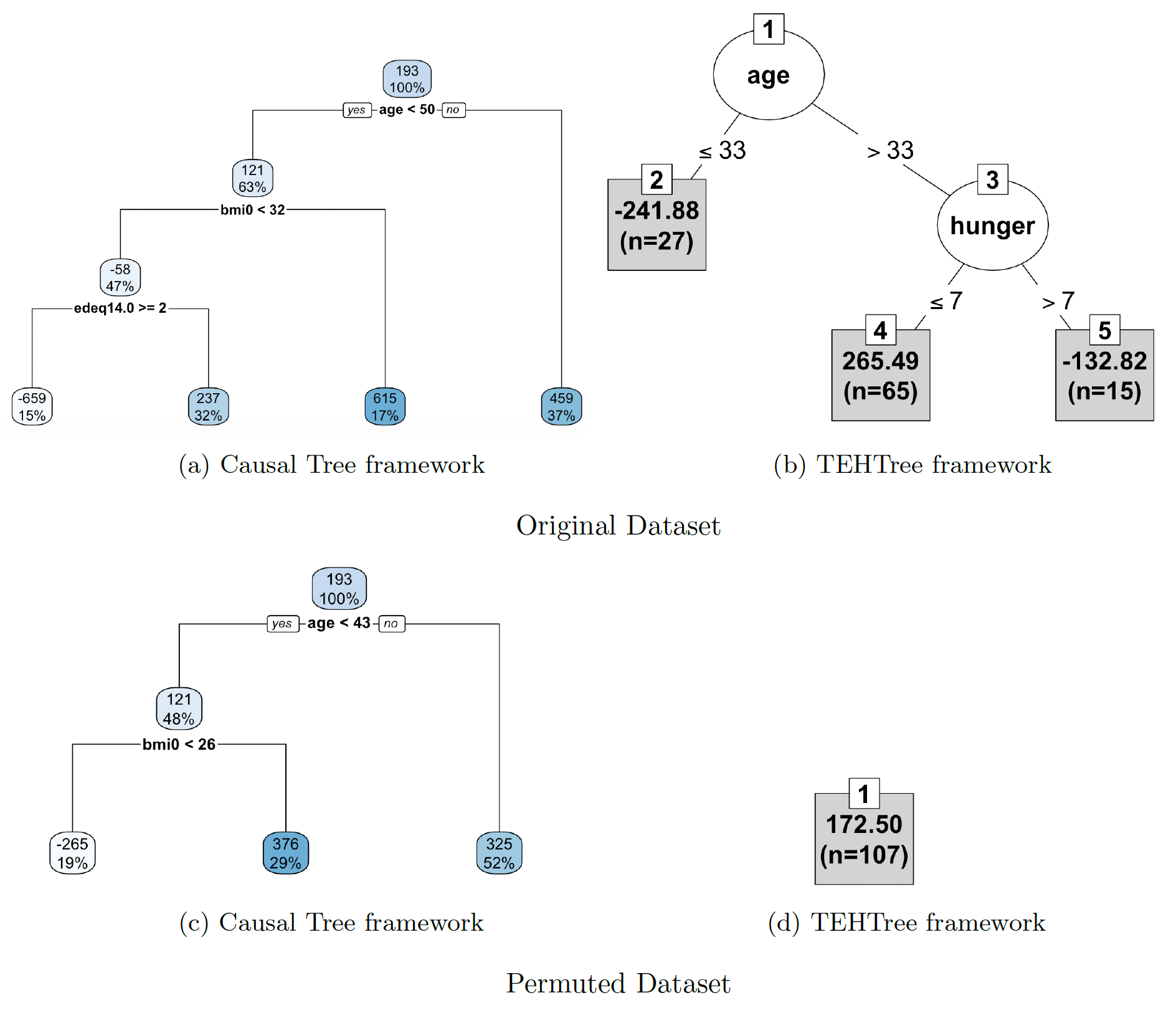}
    \caption{Results of applying Causal Tree and TEHTrees to identify heterogeneity in the effect of the intervention on daily caloric intake using data (with four different covariates) from the Box Lunch Study. Sub-figures \ref{fig:permortree}(a) and \ref{fig:permortree}(b) show the trees produced by applying the methods to the original dataset, and sub-figures \ref{fig:permortree}(c) and \ref{fig:permortree}(d) show the trees when the methods are applied to a dataset where the rows permuted to remove covariate-outcome associations.}
    \label{fig:permortree}
\end{figure}

Next, we compared the results of applying TEHTree and Causal Tree to the data. Figures \ref{fig:permortree}(a)-(b) show the TEHTree and Causal Tree that result from analyzing the BLS data. TEHTree was implemented using the modifications specified in Section \ref{sect:estimation}, and the double-sample Causal Tree method used the default parameters with the exception of setting a minimum size for splitting to 10 treated and 10 controls. While both trees identify covariates that induce heterogeneity, the covariate identified and their split points are quite different. Figures \ref{fig:permortree}(c)-(d) represent the results of applying Causal Tree and TEHTree to a permuted version of the dataset in which the rows of the covariate matrix were permuted to remove associations between covariates and the outcomes but retain within-covariate correlation. In these data, there should be no association between covariates and treatment effects, however, Causal Tree identifies three heterogeneous subgroups defined by age and BMI. TEHTree does not generate any splits, correctly reflecting the true lack of heterogeneity in the permuted data.

\section{Discussion}

Characterizing treatment effect heterogeneity is becoming a common target of secondary analyses of data from randomized controlled trials. As an alternative to methods that require that the nature of potential subgroups be pre-specified (e.g., via covariate interactions with treatment), several methods have been recently proposed to detect treatment effect heterogeneity in a more data-driven manner \citep{athey2016, Wager2017, Powers2018, kunzel2019}. However, while most of these methods incorporate procedures for preventing overfitting, they do not offer any guarantees about Type I error, i.e., the probability of identifying heterogeneous subgroups in the absence of treatment effect heterogeneity. Particularly in the context of randomized trials, explicit control of Type I error may increase the willingness of researchers to apply treatment effect heterogeneity techniques. In this paper, we propose TEHTrees, a novel method that uses a conditional inference tree framework to characterize effect heterogeneity of a binary treatment while controlling the Type I error rate. 

As shown in our simulation study, existing methods often yield much higher than nominal Type I error rates, with this error rate generally increasing with sample size. In contrast, TEHTree maintains the specified Type I error rate across most scenarios. In some scenarios, the power of TEHTrees to detect true heterogeneity is competitive with Causal Tree; in other scenarios, TEHTrees has lower power, but these discrepancies mostly arise in scenarios where Causal Tree has very high Type I error rates, i.e., its power curve lies above that of TEHTrees for both null and alternative hypotheses. Causal Trees displayed lower MSE than TEHTrees in multiple scenarios, which was most likely due to greater variability in split points for TEHTree structures. We conjecture that this variability can be attributed to the bias introduced by the matching estimator. Decreasing bias in the matching estimator, or using an alternative approach to estimating the outcomes that are used as inputs in the conditional inference tree of TEHTrees, may improve estimation of treatment effects with TEHTrees when there are continuous covariates.

TEHTrees offer a flexible approach to detecting effect heterogeneity, and its various building blocks allow for numerous modifications including the choice of a different matching algorithm, an alternative prognostic score model, the utilisation of other criteria to select the splitting variable or its split point, or the implementation of a different estimation technique. For example, the Bonferroni correction method used to find the splitting variable in TEHTrees is likely too conservative to detect small treatment effects when there are a large number of covariates in the study. Alternative multiple comparison adjustment methods should be explored in such cases, e.g., controlling the false discovery rate may be a desirable alternative. In addition, the TEHTree approach could be modified to use other regression models for determining splitting variables; indeed, any technique that tests the relevant marginal null hypotheses while accounting for the matching-induced correlation could be easily incorporated in the TEHTree framework. Our method also inherits the desirable properties of conditional inference trees including the fact that unlike the CART-based Causal Tree, TEHTrees do not favor the inclusion of continuous variables with many potential split points over categorical variables with fewer \citep{song2004study}. 

Other modifications and extensions could also improve the robustness of TEHTrees. Like many matching-based approaches, our method does not account for the variability introduced by the matching estimator, so an extra step may be required to control for the inflation of Type I error that might occur in situations when good matches are difficult to obtain due to lack of overlap in covariate distributions between treatment groups. However, in randomized studies covariates are, on average, balanced between treatment groups and hence lack of overlap between the supports of the covariate distributions is unlikely to be a problem. TEHTree's performance depends on the accuracy of the prognostic score model. As shown in the simulation, using an ensemble approach to estimating the prognostic score provides a degree of (but not total) robustness against prognostic score misspecification. One type of misspecification that we did not consider in the simulation was omission of predictors of the outcome; as with most matching-based methods, omission of such variables will decrease the quality of the resulting matches and hence lead to poorer performance. %A larger variety of scenarios may also need to be considered; this can include an increase in the complexity of the prognostic score model or to explore alternative effect sizes and treatment effect patterns (such as continuous interactions with treatment). %The prognostic score model, for example, was quite simple. 

Though we assume treatment is randomized throughout this paper, TEHTrees could be extended for use with observational data involving non-randomized treatments or exposures. However, additional assumptions and modifications to the method would be required to achieve covariate balance and ensure unbiased estimation of treatment effects.

\section*{Acknowledgments}

The authors are grateful to Dr. Simone French for permitting use of the Box Lunch Study data.

{\it Conflict of Interest}: None declared.

\bibliographystyle{unsrtnat}
\bibliography{main}

\appendix

%  To get the journal style of heading for an appendix, mimic the following.

\section{}

\subsection{Appendix A: Proofs}

\subsubsection*{Proof of Equation \ref{eq:cate-delta}}

\begin{eqnarray*}
    E(Y_i - Y_j | \mathbf{X}_i &=& \mathbf{x}, \mathbf{X}_j = \mathbf{x}', d_{ij} = d, Z_i = 1, Z_j = 0) \\&=& E(Y_{1i} - Y_{0j} | \mathbf{X}_i=\mathbf{x}, \mathbf{X}_j=\mathbf{x}', d_{ij} = d)\\
    &=& E(Y_{1i} - Y_{0i} | \mathbf{X}_i=\mathbf{x}, d_{ij} = d)
    + E(Y_{0i} - Y_{0j} | \mathbf{X}_i=\mathbf{x}, \mathbf{X}_j=\mathbf{x}', d_{ij} = d)\\
    &=& E(Y_{1i} - Y_{0i} | \mathbf{X}_i=\mathbf{x}) + \Delta^0_{ij}(\mathbf{x}, \mathbf{x}', d)\\
    &=& CATE(\mathbf{x}) + \Delta^0_{ij}(\mathbf{x}, \mathbf{x}', d)
\end{eqnarray*}

\subsubsection*{Proof of Equation \ref{eq:cate}}

\begin{eqnarray*}
E(Y_i - Y_j \mid \mathbf{X}_i &=& \mathbf{x}, \mathbf{X}_j = \mathbf{x}', \phi, Z_i = 1, Z_j = 0)\\ &=& E(Y_{1i} - Y_{0j} | \mathbf{X}_i = \mathbf{x}, \mathbf{X}_j = \mathbf{x}', \phi)\\
	&=& E(Y_{1i} - Y_{0i} | \mathbf{X}_i = \mathbf{x}, \phi) + E(Y_{0i} - Y_{0j} | \mathbf{X}_i = \mathbf{x}, \mathbf{X}_j = \mathbf{x}', \phi)\\
    &=& E(Y_{1i} - Y_{0i} | \mathbf{X}_i = \mathbf{x}) + E({Y}_{0i} - Y_{0j} | \phi)\\
    &=& E(Y_{1i} - Y_{0i} | \mathbf{X}_i = \mathbf{x}) + 0 = CATE(\mathbf{x})\\
\end{eqnarray*}

\subsection{Appendix B: Simulation Scenarios and Additional Results}
%\AVnote{All original tables can be found in simulation.tex}
%and let $\gamma$ equal either 0 (meaning there is no treatment effect heterogeneity) or 3 (meaning there is treatment effect heterogeneity).

Continuous covariates were generated from multivariate normal distributions with mean zero, unit variance, and varying pairwise correlations. Binary covariates were generated as independent Binomial$(1, 0.5)$. Continuous outcomes were generated as independent $\mathcal{N}$($\mu, 1$) with linear predictor $\mu$ as defined below in Table \ref{tab:sim_parameters}. We set $\alpha$ to 0.8, $\theta$ to 0.8 and $\beta$ to $(1.0, 0.8, 0.6, 0.4, 0.2)^T$ for the first five covariates (0 otherwise). The following models (M1 - M10) and their relevant parameters are used over multiple data scenarios, where data are simulated over $N$ different sample sizes (i.e., $N = 100, 200, 500, 1000$ and $2000$). %For continuous covariates, potential split points were tested at 5\% quantile increments.

 \begin{table}[ht]
    \centering
     \caption{Simulation parameters}
    \label{tab:sim_parameters}
    \begin{tabular}{lcl}
    \hline
    Models & (M1) & $\mu = 0.8 + 0.8 Z + \beta \mathbf{X}$\\
          & (M2) & $\mu = 0.8 + 0.8 Z + \beta \mathbf{X} + \phi I(\mathbf{X} > 0)$\\
          & (M3) & $\mu = 0.8 + 0.8 Z + \beta \mathbf{X} + \gamma Z \cdot I(X_1 > 0)$\\
          & (M4) & $\mu = 0.8 + 0.8 Z + \beta \mathbf{X} + \gamma Z \cdot X_1$\\
          & (M5) & $\mu = 0.8 + 0.8 Z + \beta \mathbf{X} + \gamma_1 Z \cdot X_1 + \gamma_2 Z \cdot I(X_1 > 0)$\\
          & (M6) & $\mu = 0.8 + 0.8 Z + \beta \mathbf{X} + \gamma Z \cdot I(-0.5 < X_1 < 0.5)$\\
          & (M7) & $\mu = 0.8 + 0.8 Z + \beta \mathbf{X} + \gamma Z \cdot \sin(\eta X_1)$\\
          & (M8) & $\mu = 0.8 + 0.8 Z + \beta \mathbf{X} + \gamma_1 Z \cdot I(X_1 > 0) + \gamma_2 Z \cdot I(X_2 > 0))$\\
          & (M9) & $\mu = 0.8 + 0.8 Z + \beta \mathbf{X} + \gamma_1 Z \cdot X_1 + \gamma_2 Z \cdot I(X_2 > 0)$\\
          & (M10) & $\mu = 0.8 + 0.8 Z + \beta \mathbf{X} + \gamma Z \cdot\mathbf{X}$\\
          & (M11) & $\mu = 0.8 + 0.8 Z + \beta \mathbf{X} + \gamma_1 Z \cdot \mathbf{X}_C + \gamma_2 Z \cdot \mathbf{X}_B$\\
          &      & $\beta = (1.0, 0.8, 0.6, 0.4, 0.2, 0, \dots, 0)^T$\\
          
    \hline
    Covariates & (C1) &  $\mathbf{X}$ = 5 independent binary covariates\\
             &   (C2) & $\mathbf{X}$ = 5 independent continuous covariates\\
              & (C3) & $\mathbf{X}$ = 10 independent continuous covariates\\
    \hline
    Coefficients & (P1) & $\phi_1 = 3,  \phi_2 = \phi_3 = \phi_4 = \phi_5 = 0$\\
            & (P2) & $\phi_1 = 1, \phi_2 = \phi_3 = \phi_4 = \phi_5 = 0$\\
            & (P3) & $\phi_1 = \phi_2 = 1, \phi_3 = \phi_4 = \phi_5 =0$\\
            & (P4) & $\gamma = 1$\\
            & (P5) & (i) $\gamma = 2$, (ii) $\gamma = 1$\\
            & (P6) & (i) $\gamma_1 = \gamma_2 = 1$, (ii) $\gamma_1 = 1, \gamma_2 = -1$,\\\
           & &(iii) $\gamma_1 = -1, \gamma_2 = 1$, (iv) $\gamma_1 = \gamma_2 = -1$\\
            & (P7) & (i) $\gamma = 3$, (ii) $\gamma = 2$, (iii) $\gamma = 1$\\
            & (P8) & (i) $\gamma = 2, \eta = 2$, (ii) $\gamma = 1, \eta = 2$,\\\
            & &(iii) $\gamma = 2, \eta = 1.5$, (iv) $\gamma = 1, \eta = 1.5$\\
            & (P9) & (i) $\gamma_1 = 3, \gamma_2 = 3$, (ii) $\gamma_1 = 1, \gamma_2 = 1$,\\\
            & & (iii) $\gamma_1 = 3, \gamma_2 = -3$, (iv) $\gamma_1 = 1, \gamma_2 -3$\\
            & (P10) & (i) $\gamma_1 = \gamma_2 = 1$, (ii) $\gamma_1 = 1, \gamma_2 = -1$,\\
            & & (iii) $\gamma_1 = -1, \gamma_2 = 1$, (iv) $\gamma_1 = \gamma_2 = -1$\\
            & (P11) & (i) $\gamma = 1$, (ii) $\gamma = 2$, (iii) $\gamma = 6$
    \end{tabular}
\end{table}

\begin{table}[htb!]
    \centering
    \caption{Simulation scenarios corresponding to main figures. Scenario codes are fully described in Table \ref{tab:sim_parameters}.}\label{tab:sim_scenarios}
    \begin{tabular}{ll}
    \hline
    \multicolumn{2}{l}{\textbf{Evaluating Type I Error}}\\
    \hline
    \multicolumn{2}{l}{\emph{Figure \ref{fig:plots1}(a)}}\\
    (M1) (C1)-(C3) & Correctly specified prognostic score\\
    \hline
    \multicolumn{2}{l}{\emph{Figure \ref{fig:plots1}(b)}}\\
    (M2) (C1) (P1)-(P3) & Incorrectly specified prognostic score\\
    \hline
    \multicolumn{2}{l}{\textbf{Evaluating Power, MSE and Tree Characteristics}}\\
    \hline
    \multicolumn{2}{l}{\emph{Figures \ref{fig:powersone}(a)} \& \emph{\ref{fig:char}(a)}}\\
    (M3) (C1)-(C3) (P4) & Heterogeneity according to $I(X_1 > 0)$\\
    \emph{Figure \ref{fig:char}(b)} & [(i)-(iii) refers to settings (C1)-(C3)] \\
    (M3) (C2)-(C3) (P4) & [(ii)-(iii) refers to settings (C2)-(C3)]\\
    \hline
    \multicolumn{2}{l}{\emph{Figure \ref{fig:powersone}(b)} \& \emph{\ref{fig:powercomp}(a)}}\\
    (M4) (C3) (P5) (i)-(ii) & Heterogeneity according to $X_1$\\
    \hline
    \multicolumn{2}{l}{\emph{Figures \ref{fig:powersone}(c) \& \ref{fig:powersone}(f)}}\\
    (M5) (C3) (P6) (i)-(iv) & Heterogeneity according to $I(X_1 > 0)$ and $X_1$\\
    \hline
    \multicolumn{2}{l}{\emph{Figures \ref{fig:msepoweradd}(a)}, \emph{\ref{fig:powercomp}(b)} \& \emph{\ref{fig:msepoweradd}(d)}}\\
    (M6) (C3) (P7) (i)-(iii) & Heterogeneity according to $I(-0.5 < X_1 < 0.5)$\\
    \hline
    \multicolumn{2}{l}{\emph{Figures \ref{fig:powersone}(d)} \& \emph{\ref{fig:powersone}(g)}}\\
    \emph{Figures \ref{fig:char}(c)} \& \emph{\ref{fig:char}(d)} & Heterogeneity according to $\sin(\eta X_1)$\\ 
    (M7) (C3) (P8) (i)-(iv)\\
    \hline
    \multicolumn{2}{l}{\emph{Figures \ref{fig:msepoweradd}(b), {\ref{fig:powercomp}(c)}} \& \emph{\ref{fig:msepoweradd}(e)}}\\
    (M8) (C3) (P9) (i)-(iv) & Heterogeneity according to $I(X_1 > 0)$ and $I(X_2 > 0)$\\
    \hline
    \multicolumn{2}{l}{\emph{Figures \ref{fig:powersone}(e) \& \ref{fig:powersone}(h)}}\\
    (M9) (C3) (P10) (i)-(iv) & Heterogeneity according to $X_1$ and $I(X_2 > 0)$\\
    \hline
    \multicolumn{2}{l}{\emph{Figures \ref{fig:msepoweradd}(c)} \& \emph{\ref{fig:msepoweradd}(f)}}\\
    (M10) (C3) (P11) (i)-(iii) & Heterogeneity according to $X_1, \dots, X_{10}$\\
    \hline
    \multicolumn{2}{l}{\textbf{Subgroup Identification}}\\
    \hline
    \multicolumn{2}{l}{\emph{Figure \ref{fig:subgroup}}}\\
    (M1) (C2) \& (M3) (C2) & Heterogeneity according to $X_1$\\
    \multicolumn{2}{l}{\emph{Figure \ref{fig:subsplits}}}\\
    (M3) Covariates $X_1$ \& $X_2$ & Heterogeneity according to $X_1$\\
    \hline
    \multicolumn{2}{l}{\textbf{Variable Selection}}\\
    \hline
    \multicolumn{2}{l}{\emph{Figure \ref{fig:varsel}}}\\
    (M11) (C3) & Heterogeneity according to $X_1$ and $X_6$
    \end{tabular}
\end{table}

\subsection*{Additional scenarios for Models M1 - M10}

The results evaluated using Model M1 and Model M2 %Equations \ref{eq:typeeq1} and \ref{eq:typeeq2} 
were displayed in Figures \ref{fig:plots1}(a) and \ref{fig:plots1}(b) using data generated over $N$ = 100, 200, 500, 1000 and 2000. However, the results displayed in Figure \ref{fig:plots1} assumes the absence of pairwise correlation ($\rho = 0$) among the five covariates. %Model M1 %Equation \ref{eq:typeeq1} was varied for different number and type of covariates and 
Table \ref{tab:tab1} present results for non-zero values of pairwise correlation $(\rho = 0.2, \rho = 0.4, \rho = 0.6, \rho = 0.8)$ over continuous covariates (m = 5) at $N$ = 200 and $N$ = 500. Similarly, Model M2
%Equation \ref{eq:typeeq2} 
was varied using different vectors of coefficients for five continous covariates and Table \ref{tab:tab2} presents results for two additional vectors of coefficients.

\begin{table}[ht]
\centering
\caption{Type I Error rate when there is no treatment effect heterogeneity (i.e., when $\gamma = 0$) and the data are generated according to Model M1. This table presents results for non-zero values of pairwise correlation $(\rho = 0.2, \rho = 0.4, \rho = 0.6, \rho = 0.8)$ over continuous covariates (m = 5) at $N$ = 200 and $N$ = 500. 
}
\label{tab:tab1}
\begin{tabular}{cccccc}
  \hline
 Covariate Type & N & m & $\rho$ & \multicolumn{2}{c}{Type I Error Rate}  \\ 
   &  &  &  & TT & CT \\
  \hline
  Continuous & 200 & 5 & 0.2 & 0.030 & 0.687  \\ 
  Continuous & 200 & 5 & 0.4 & 0.026 & 0.767  \\ 
  Continuous & 200 & 5 & 0.6 & 0.026 & 0.825  \\ 
  Continuous & 200 & 5 & 0.8 & 0.050 & 0.851  \\ 
  Continuous & 500 & 5 & 0.2 & 0.023 & 0.970  \\ 
  Continuous & 500 & 5 & 0.4 & 0.026 & 0.991  \\ 
  Continuous & 500 & 5 & 0.6 & 0.030 & 0.994  \\ 
  Continuous & 500 & 5 & 0.8 & 0.045 & 0.998  \\  
   \hline
\end{tabular}
\end{table}

\begin{table}[ht]
\centering
\caption{Type I Error rate of TEHTrees and Causal Tree when there is no treatment effect heterogeneity and the data are generated according to Model M2.
%Equation \ref{eq:typeeq2}. 
The last column shows the Type I Error of TEHTrees when the true prognostic scores are used in matching.}
\label{tab:tab2}
\begin{tabular}{cccccccccc}
  \hline
  N & m & $\phi_1$ & $\phi_2$ & $\phi_3$ & $\phi_4$ & $\phi_5$ & \multicolumn{3}{c}{Type I Error Rate}  \\ 
    & & & &  &  &  & TT & CT & TT (true PS)  \\
  \hline
   100 & 5 & 0.5 & 1 & 0& 0& 0& 0.024 & 0.429 & 0.053 \\  
   200 & 5 & 0.5 & 1 & 0& 0& 0& 0.021 & 0.627 & 0.052  \\  
   500 & 5 & 0.5 & 1 & 0& 0& 0& 0.031 & 0.904 & 0.062  \\  
   1000 & 5 & 0.5 & 1 & 0& 0& 0& 0.056 & 0.986 & 0.048  \\ 
   2000 & 5 & 0.5 & 1 & 0& 0& 0& 0.155 & 0.995 & 0.060  \\ 
     100 & 5 & 1 & 1 & 1&1& 1& 0.026 & 0.433 & 0.053 \\  
   200 & 5 & 1 & 1 & 1&1& 1& 0.019 & 0.677 & 0.046  \\  
  500 & 5 & 1 & 1 & 1&1& 1& 0.032 & 0.956 & 0.044 \\  
   1000 & 5 & 1 & 1 & 1&1& 1& 0.092  & 0.997 & 0.046 \\  
   2000 & 5 & 1 & 1 & 1&1& 1& 0.144 & 0.998 & 0.054  \\  
   \hline
\end{tabular}
\end{table}

\begin{table}[ht]
\centering
\caption{Power of TEHTrees and Causal Trees when data are generated according to Model M3 
%Equation \ref{eq:powerdist1} 
and there is treatment effect heterogeneity.}
\label{tab3}
\begin{tabular}{cccccc}
  \hline
 Covariate Type & N & m & $\rho$ & \multicolumn{2}{c}{Power}  \\ 
    &  &  &  & TT & CT \\
  \hline
  Continuous & 200 & 5 & 0.2 & 0.68 & 0.93  \\ 
  Continuous & 200 & 5 & 0.4 & 0.72 & 0.89  \\ 
  Continuous & 200 & 5 & 0.6 & 0.73 & 0.88  \\ 
  Continuous & 200 & 5 & 0.8 & 0.66 & 0.83  \\ 
  Continuous & 500 & 5 & 0.2 & 0.97 & 0.98  \\ 
  Continuous & 500 & 5 & 0.4 & 0.98 & 0.98  \\ 
  Continuous & 500 & 5 & 0.6 & 0.98 & 0.95  \\ 
  Continuous & 500 & 5 & 0.8 & 0.95 & 0.94  \\  
   \hline
\end{tabular}
\end{table}

\begin{table}[htb!]
\centering
\caption{Characteristics of the trees including the median and mean of the first split point on $X_1$ (the variable with heterogeneous treatment effects). Data are generated using Model M3.
%Equation \ref{eq:powerdist1}
}
\label{tab11}
\begin{tabular}{cccccccccc}
  \hline
 Covariate & N & m & $\rho$ & \multicolumn{2}{c}{Median}  & \multicolumn{2}{c}{Mean}  \\ 
type &  &  &  & \multicolumn{2}{c}{split point}  & \multicolumn{2}{c}{split point}  \\ 
   &  &  &  & TT & CT & TT & CT \\

  \hline
Continuous & 100 & 5 & 0.0 &  -0.01 & -0.01 & -0.01 & -0.01   \\ 
  Continuous & 200 & 5 & 0.0 & 0.00 & 0.00 & 0.00 & 0.00  \\ 
  Continuous & 500 & 5 & 0.0 & 0.00 & 0.00 & 0.00 & 0.00   \\ 
  Continuous & 1000 & 5 & 0.0 &  0.00 & -0.00 & 0.00 & -0.00  \\ 
  Continuous & 2000 & 5 & 0.0 &  0.00 & -0.00 & 0.00 & -0.00   \\ 
  Continuous & 100 & 10 & 0.0 &  -0.01 & -0.03 & -0.03 & -0.02   \\ 
  Continuous & 200 & 10 & 0.0 & 0.00 & 0.01 & 0.02 & 0.01   \\
  Continuous & 500 & 10 & 0.0 & 0.00 & 0.00 & 0.00 & 0.00 \\
  Continuous & 1000 & 10 & 0.0 &  0.00 & 0.00 & 0.00 & 0.00  \\ 
  Continuous & 2000 & 10 & 0.0 & 0.00 & 0.00 & 0.00 & -0.00  \\ 
  Continuous & 200 & 5 & 0.2 & 0.00 & 0.00 & 0.00 & 0.00   \\ 
  Continuous & 200 & 5 & 0.4 &  0.00 & -0.00 & 0.00 & -0.00   \\ 
  Continuous & 200 & 5 & 0.6 & 0.00 & 0.00 & -0.01 & -0.00   \\ 
  Continuous & 200 & 5 & 0.8 &  0.00 & 0.00 & 0.01 & -0.01  \\ 
  Continuous & 500 & 5 & 0.2 &  0.00 & -0.00 & 0.00 & -0.00  \\ 
  Continuous & 500 & 5 & 0.4 &  0.00 & -0.00 & -0.01 & -0.00   \\ 
  Continuous & 500 & 5 & 0.6 & -0.01 & -0.00 & 0.00 & -0.00   \\ 
  Continuous & 500 & 5 & 0.8 &  -0.01 & -0.00 & -0.01 & -0.01  \\ 
   \hline
\end{tabular}
\end{table}

\begin{table}[htb!]
\centering
\caption{Power and MSE of the estimated average treatment effect for TEHTrees (TT) and Causal Tree (CT) when data are generated according to Model M4. %Equation \ref{eq:powercont1}. 
 Power is defined to be the probability of the tree making a split on $X_1$. MSE is also computed for Causal Forests (CF) using the estimated individual treatment effects against the generated `true' treatment effects.} %Equation \ref{eq:powercont1}.}
\label{tab6}
\begin{tabular}{cccccccccccccc}
  \hline
 $\gamma$ & N  & $\rho$ & \multicolumn{2}{c}{Power} & \multicolumn{2}{c}{\# nodes}  &  \multicolumn{2}{c}{\% non-$X_1$ split} & \multicolumn{3}{c}{MSE} \\ 
    &  &    & TT & CT  & TT & CT & TT & CT & TT & CT & CF\\
  \hline
  1 & 200 & 0.2 & 0.70 & 0.83 & 2.15 & 5.09 &  0.05 & 0.87 &  0.89 & 1.22 & 0.64\\ 
  1 & 500 & 0.2 & 0.99 & 0.99 & 2.90 & 11.53 &  0.09 & 0.99 &   0.50 & 0.80 & 0.27\\ 
  1 & 200 & 0.4 & 0.76 & 0.80 & 2.20 & 5.18 &  0.09 & 0.90  &  0.93 & 1.28 & 0.58\\ 
  1 & 500 & 0.4 & 1.00 & 0.98 & 2.99 & 11.60 &  0.13 & 0.99 &  0.52 & 0.85 & 0.26\\ 
  1 & 200 & 0.6 & 0.74 & 0.76 & 2.25 & 5.22 &  0.17 & 0.91 & 0.97 & 1.26 & 0.51\\ 
  1 & 500 & 0.6 & 0.99 & 0.96 & 3.10 & 11.11 &  0.19 & 0.97 & 0.51 & 0.82 & 0.24\\ 
  1 & 200 & 0.8 & 0.64 & 0.71 & 2.24 & 5.21 &  0.29 & 0.89 & 1.01 & 1.09 & 0.43\\ 
  1 & 500 & 0.8 & 0.96 & 0.93 & 3.20 & 10.88 &  0.33 & 0.96 &  0.52 & 0.70 & 0.20\\ 
   \hline
\end{tabular}
\end{table}

\newpage 

\begin{table}[htb!]
\centering
\caption{Power and MSE of the estimated average treatment effect for TEHTrees (TT) and Causal Tree (CT) when data are generated according to Model M5. %Equation \ref{eq:powercont3}. 
Power is defined to be the probability of the tree making a split on $X_1$. MSE is also computed for Causal Forests (CF) using the estimated individual treatment effects against the generated `true' treatment effects.}%Equation \ref{eq:powercont3}.}
\label{tab8}
\begin{tabular}{p{0.35cm}p{0.35cm}ccccccccccc}
  \hline
 $\gamma_1$ & $\gamma_2$ & N & $\rho$ & \multicolumn{2}{c}{Power} & \multicolumn{2}{c}{\# nodes}  &  \multicolumn{2}{c}{\% non-$X_1$ split} & \multicolumn{3}{c}{MSE}  \\ 
    &  &  &  & TT & CT  & TT & CT & TT & CT & TT & CT & CF\\
  \hline
  -1 & -1 & 200 & 0.2 & 0.82 & 0.95 & 2.14 & 4.42 &  0.06 & 0.68 & 1.14 & 1.04 \\ 
  -1 & -1 & 500 & 0.2 & 1.00 & 1.00 & 2.71 & 10.46 &  0.15 & 0.92 &  0.59 & 0.69 \\ 
  -1 & -1 & 200 & 0.4 & 0.81 & 0.98 & 2.19 & 4.95 &  0.13 & 0.82 & 1.16 & 1.06 \\ 
  -1 & -1 & 500 & 0.4 & 1.00 & 1.00 & 2.88 & 11.36 &  0.29 & 0.99& 0.59 & 0.70 \\ 
  -1 & -1 & 200 & 0.6 & 0.81 & 0.96 & 2.26 & 5.03 &  0.23 & 0.86 & 1.14 & 1.04 \\ 
  -1 & -1 & 500 & 0.6 & 1.00 & 1.00 & 3.16 & 11.32 &  0.44 & 0.98 & 0.58 & 0.66\\ 
  -1 & -1 & 200 & 0.8 & 0.68 & 0.93 & 2.37 & 5.03 &  0.42 & 0.91 & 1.09 & 0.92 \\ 
  -1 & -1 & 500 & 0.8 & 0.96 & 1.00 & 3.44 & 11.01 &  0.60 & 0.98 & 0.57 & 0.59\\ 
   \hline
\end{tabular}
\end{table}

\begin{table}[htb!]
\centering
\caption{Power and MSE of the estimated average treatment effect for TEHTrees and Causal Tree when data are generated according to Model M6. %Equation \ref{eq:powerdist2}. 
Power is defined to be the probability of the tree making a split on $X_1$. MSE is also computed for Causal Forests (CF) using the estimated individual treatment effects against the generated `true' treatment effects. %Equation \ref{eq:powerdist2}.
}
\label{tab4}
\begin{tabular}{ccccccccccccccc}
  \hline
 $\gamma$ & N  & $\rho$ & \multicolumn{2}{c}{Power} & \multicolumn{2}{c}{\# nodes}  & \multicolumn{2}{c}{\% non-$X_1$ split} & N & \multicolumn{3}{c}{MSE} \\ 
    &  &    & TT & CT  & TT & CT & TT & CT & &  TT & CT & CF\\
  \hline
  3 & 200 & 0.2 & 0.13 & 1.00 & 2.88 & 19.59 &  0.17 & 0.99 & 1000 &  1.97 & 0.57 & 0.20\\ 
  3 & 500 & 0.2 & 0.31 & 1.00 & 3.48 & 38.13 &  0.40 & 1.00 & 2000 &  1.64 & 0.43 & 0.09\\ 
  3 & 200 & 0.4 & 0.13 & 1.00 & 2.83 & 21.14 &  0.21 & 1.00 & 1000 &  1.98 & 0.60 & 0.19\\ 
  3 & 500 & 0.4 & 0.35 & 1.00 & 3.60 & 38.00 & 0.47 & 1.00 & 2000 &  1.66 & 0.42 & 0.09\\ 
  3 & 200 & 0.6 & 0.15 & 0.99 & 2.85 & 20.30 &  0.28 & 0.99 & 1000 &  1.97 & 0.58 & 0.18\\ 
  3 & 500 & 0.6 & 0.36 & 1.00 & 3.63 & 36.76 &  0.53 & 1.00 & 2000 &  1.70 & 0.39 & 0.09\\ 
  3 & 200 & 0.8 & 0.17 & 0.98 & 2.97 & 19.36 &  0.32 & 0.98 & 1000 &  1.97 & 0.54 & 0.17\\ 
  3 & 500 & 0.8 & 0.32 & 1.00 & 3.42 & 33.33 &  0.59 & 1.00 & 2000 &  1.75 & 0.33 & 0.08\\ 
   \hline
\end{tabular}
\end{table}

\begin{table}[ht]
\centering
\caption{Power and MSE of the estimated average treatment effect for TEHTrees (TT) and Causal Tree (CT) when data are generated according to Model M7. %Equation \ref{eq:powercont2}. 
Power is defined to be the probability of the tree making a split on $X_1$. }%Equation \ref{eq:powercont2}}
\label{tab7}
\begin{tabular}{p{0.35cm}p{0.35cm}cccccccccc}
  \hline
 $\gamma$ & N & $\eta$ & $\rho$ & \multicolumn{2}{c}{Power} & \multicolumn{2}{c}{\# nodes}  &  \multicolumn{2}{c}{\% non-$X_1$ split} & \multicolumn{2}{c}{MSE}   \\ 
    &  &  &  & TT & CT  & TT & CT &  TT & CT & TT & CT \\
  \hline
  2 & 200 & 1.5 & 0.2 & 0.40 & 0.96 & 2.10 & 4.90 & 0.04 & 1.62 &  1.95 & 1.82  \\ 
  2 & 500 & 1.5 & 0.2 & 0.77 & 1.00 & 2.23 & 11.20 &  0.06 & 0.97 & 1.42 & 1.37 \\  
  2 & 200 & 1.5 & 0.4 & 0.48 & 0.94 & 2.10 & 5.14 &  0.04 & 0.87 &  1.94 & 1.87 \\ 
  2 & 500 & 1.5 & 0.4 & 0.86 & 0.99 & 2.24 & 11.52 &  0.08 & 0.98 & 1.38 & 1.40\\ 
  2 & 200 & 1.5 & 0.6 & 0.49 & 0.92 & 2.11 & 5.20 &  0.09 & 0.88 &  1.98 & 1.86\\ 
  2 & 500 & 1.5 & 0.6 & 0.87 & 0.99 & 2.26 & 11.24 &  0.10 & 0.98&  1.38 & 1.35 \\ 
  2 & 200 & 1.5 & 0.8 & 0.47 & 0.86 & 2.15 & 5.13 & 0.16 & 0.88 & 2.05 & 1.80\\ 
  2 & 500 & 1.5 & 0.8 & 0.83 & 0.98 & 2.30 & 10.82 & 0.17 & 0.97 &  1.43 & 1.26 \\
  \hline
\end{tabular}
\end{table}

\begin{table}[htb!]
\centering
\caption{Power and MSE of the estimated average treatment effect using TEHTrees (TT) and Causal Trees (CT) when data are generated according to Model M8. % Equation \ref{eq:powerdist3}. 
Power is defined to be the probability of making a split on $X_1$ or $X_1$ (``Power Any'') or on $X_1$ and $X_2$ (``Power All'')} %Equation \ref{eq:powerdist3}.}
\label{tab5}
\begin{tabular}{p{0.35cm}p{0.35cm}ccccccccccc}
  \hline
 $\gamma_1$ & $\gamma_2$ & N & $\rho$ & \multicolumn{2}{c}{Power Any} & \multicolumn{2}{c}{Power All} & \multicolumn{2}{c}{\# nodes}  &  N & \multicolumn{2}{c}{MSE} \\ 
    &  &  &  & TT & CT & TT & CT & TT & CT &  & TT & CT   \\
  \hline
  3 & 3 & 200 & 0.2 & 0.75 & 0.94 & 0.28 & 0.78 & 2.49 & 5.01 &  1000 & 4.83 & 5.23 \\ 
  3 & 3 & 500 & 0.2 & 0.96 & 0.99 & 0.92 & 0.97 & 3.91 & 9.87 &  2000 & 4.59 & 4.93 \\ 
  3 & 3 & 200 & 0.4 & 0.88 & 0.92 & 0.31 & 0.71 & 2.56 & 4.99 &   1000 & 4.89 & 5.32\\ 
  3 & 3 & 500 & 0.4 & 1.00 & 0.98 & 0.95 & 0.97 & 3.92 & 10.21 &   2000 & 4.56 & 4.92 \\ 
  3 & 3 & 200 & 0.6 & 0.94 & 0.91 & 0.28 & 0.64 & 2.57 & 4.92 &  1000 & 4.99 & 5.27\\ 
  3 & 3 & 500 & 0.6 & 1.00 & 0.97 & 0.91 & 0.96 & 3.83 & 10.11 &   2000 & 4.52 & 4.86\\ 
  3 & 3 & 200 & 0.8 & 0.92 & 0.91 & 0.21 & 0.58 & 2.58 & 4.97 & 1000 & 5.02 & 5.18\\ 
  3 & 3 & 500 & 0.8 & 1.00 & 0.98 & 0.71 & 0.95 & 3.52 & 9.87 & 2000 & 4.58 & 4.82 \\ 
   \hline
\end{tabular}
\end{table}

\begin{figure}[htb!]
    \centering
    \includegraphics[scale = 0.5]{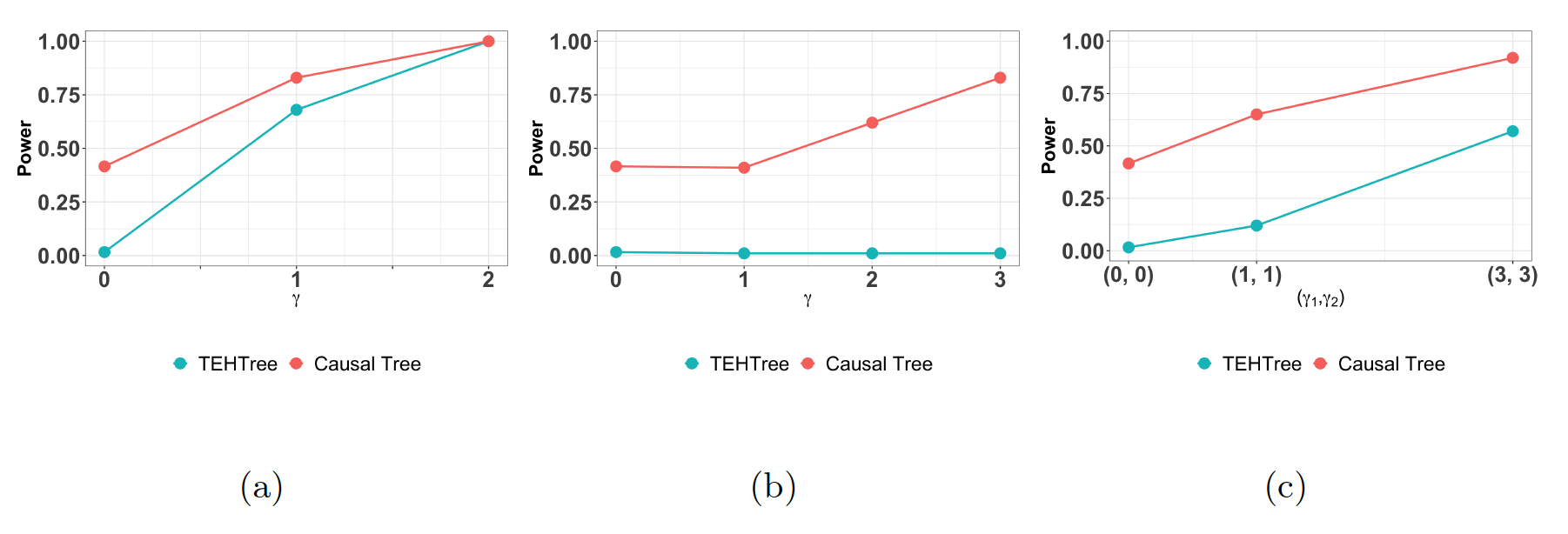}
    \caption{Power of TEHTrees and Causal Trees at different levels of $\gamma$, using scenarios for Models M4, M6 and M8 and when N = 200.}
    \label{fig:powercomp}
\end{figure}

\begin{table}[htb!]
\centering
\caption{Power and MSE of the estimated average treatment effect for TEHTrees (TT) and Causal Tree (CT) when data are generated according to Model M9. 
%Equation \ref{eq:powercont4}. 
Power is defined to be the probability of making a split on $X_1$ or $X_1$ (``Power Any'') or on $X_1$ and $X_2$ (``Power All''). }%Equation \ref{eq:powercont4}.}
\label{tab9}
\begin{tabular}{p{0.35cm}p{0.35cm}cccccccccccc}
  \hline
 $\gamma_1$ & $\gamma_2$ & N & $\rho$ & \multicolumn{2}{c}{Power Any} & \multicolumn{2}{c}{Power All} & \multicolumn{2}{c}{\# nodes}  &   \multicolumn{2}{c}{MSE} \\ 
    &  &  &  & TT & CT & TT & CT & TT & CT &  TT & CT \\
  \hline
  -1 & -1 & 200 & 0.2 & 0.60 & 0.93 & 0.04 & 0.48 & 2.17 & 4.69 &   1.19 & 1.11\\ 
  -1 & -1 & 500 & 0.2 & 0.99 & 1.00 & 0.40 & 0.90 & 3.17 & 10.69 &  0.77 & 0.76  \\ 
  -1 & -1 & 200 & 0.4 & 0.75 & 0.97 & 0.05 & 0.55 & 2.22 & 5.04 &    1.21 & 1.15 \\ 
  -1 & -1 & 500 & 0.4 & 1.00 & 1.00 & 0.49 & 0.94 & 3.51 & 11.64 &  0.77 & 0.79  \\ 
  -1 & -1 & 200 & 0.6 & 0.79 & 0.96 & 0.07 & 0.50 & 2.29 & 5.07 &    1.21 & 1.10\\ 
  -1 & -1 & 500 & 0.6 & 0.99 & 1.00 & 0.52 & 0.93 & 3.75 & 11.77 &    0.74 & 0.76\\ 
  -1 & -1 & 200 & 0.8 & 0.74 & 0.93 & 0.08 & 0.46 & 2.39 & 5.02 &    1.11 & 0.98\\ 
  -1 & -1 & 500 & 0.8 & 0.97 & 1.00 & 0.42 & 0.88 & 3.83 & 10.95 &   0.66 & 0.64\\ 
   \hline
\end{tabular}
\end{table}

\begin{table}[htb!]
\centering
\caption{Power and MSE of the estimated average treatment effect for TEHTrees (TT) and Causal Tree (CT) when data are generated according to Model M10. %Equation \ref{eq:powercont5}. 
Power is defined to be the probability of the tree making a split on any  variables (``Power Any'') or on all variables with treatment effect heterogeneity (``Power All''). $\gamma$ is the coefficient of the interaction term. }%MSE of the estimated average treatment effect using TEHTrees (TT) and Causal Tree (CT) when data are generated according to Equation \ref{eq:powercont5}.}
\label{tab10}
\begin{tabular}{p{0.35cm}p{0.35cm}cccccccccc}
  \hline
 $\gamma$ & Vars & N & $\rho$  & \multicolumn{2}{c}{Power Any} & \multicolumn{2}{c}{Power All} & \multicolumn{2}{c}{\# nodes}  &   \multicolumn{2}{c}{MSE}  \\ 
    & int.  &  &  & TT & CT & TT & CT & TT & CT &  TT & CT \\
  \hline
%   0 & 100 & 0 &  & 0. & 0.232 & 0. & 0.034  \\ 
%   0 & 200 & 0 &  & 0. & 0.399 & 0. & 0.167  \\ 
%   0 & 500 & 0 &  & 0. & 0.775 & 0. & 0.535  \\ 
%   0 & 1000 & 0 &  & 0. & 0.954 & 0. & 0.848  \\ 
%   0 & 2000 & 0 &  & 0. & 0.988 & 0. & 0.964  \\ 
  6 & 12 & 200 & 0.2 & 0.15 & 0.47 & 0.04 & 0.37 & 2.32 & 5.87 &   37.45 & 34.62  \\ 
  6 & 12 & 500 & 0.2 & 0.39 & 0.88 & 0.12 & 0.86 & 2.84 & 14.01 &  36.23 & 24.69 \\ 
  6 & 12 & 200 & 0.4 & 0.16 & 0.67 & 0.04 & 0.50 & 2.39 & 5.62  &  41.59 & 36.66\\ 
  6 & 12 & 500 & 0.4 & 0.42 & 0.98 & 0.12 & 0.97 & 2.83 & 14.77  &   39.48 & 23.26\\ 
  6 & 12 & 200 & 0.6 & 0.20 & 0.89 & 0.05 & 0.70 & 2.53 & 6.17  &   47.96 & 35.62\\ 
  6 & 12 & 500 & 0.6 & 0.43 & 1.00 & 0.12 & 1.00 & 2.90 & 15.35  &   43.91 & 22.28 \\ 
  6 & 12 & 200 & 0.8 & 0.23 & 0.98 & 0.06 & 0.78 & 2.75 & 6.47  &  55.66 & 32.80\\ 
  6 & 12 & 500 & 0.8 & 0.43 & 1.00 & 0.14 & 1.00 & 3.16 & 14.93  &  47.57 & 20.80  \\ 
  \hline
\end{tabular}
\end{table}

\begin{figure}[htb!]
    \centering
    \includegraphics[scale = 0.5]{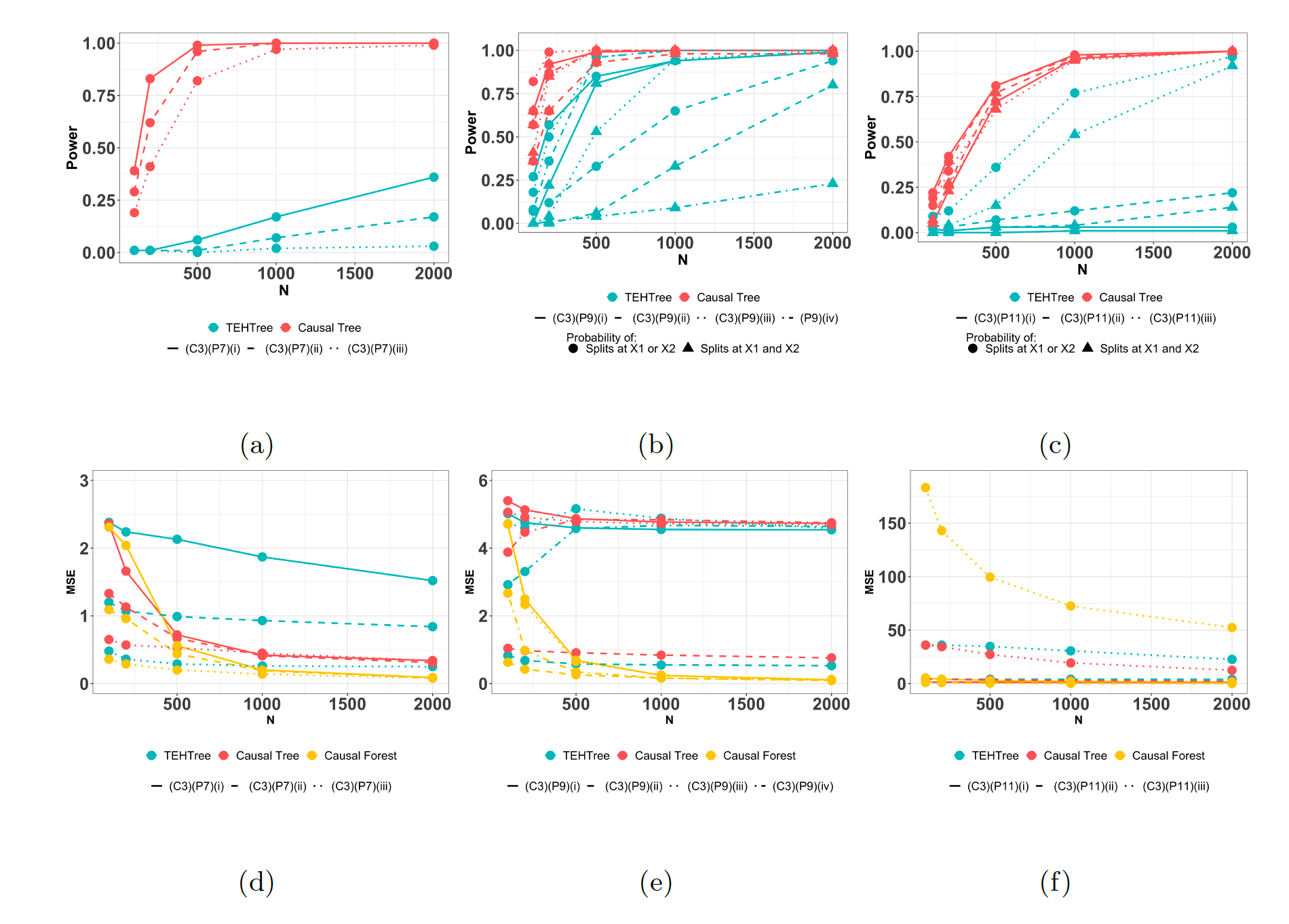}
    \caption{Power (Figure \ref{fig:msepoweradd}(a)-(c)) and Mean-squared error (MSE) (Figure \ref{fig:msepoweradd}(d)-(f)) of the estimated average treatment effect. The data are generated using Models M6, M8 and M10.}
    \label{fig:msepoweradd}
\end{figure}

\begin{figure}[ht]
    \centering
    \includegraphics[scale = 0.5]{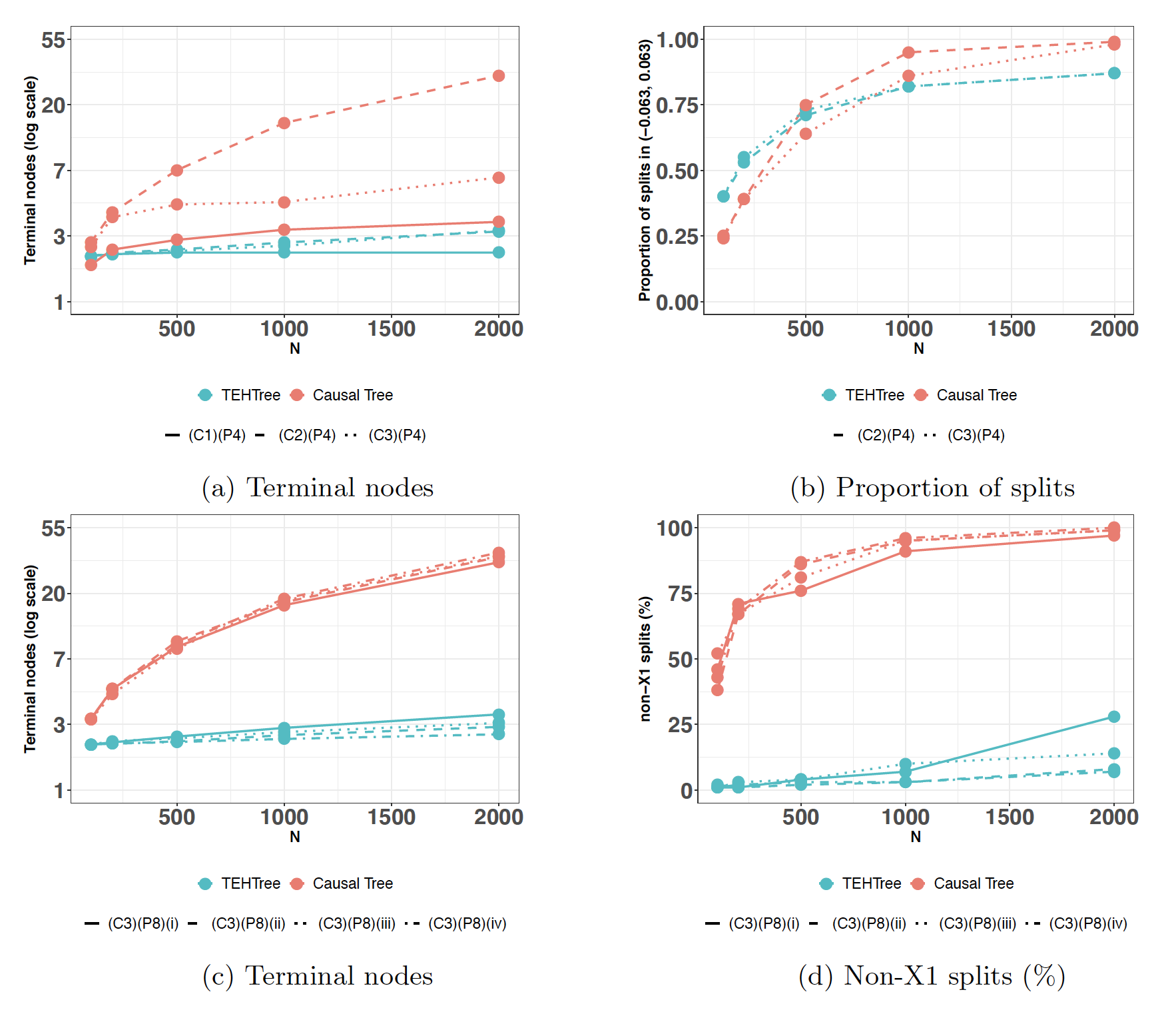}
    \caption{Characteristics of the trees built by TEHTrees (TT) and Causal Trees (CT) when heterogeneity is due to $I(X_1 > 0)$ (in Model M3) and the $\sin(\eta X_1)$ term (in Model M7). Figure \ref{fig:char}(a) and Figure \ref{fig:char}(b) correspond to data generated using Model M3 and Figure \ref{fig:char}(c) and Figure \ref{fig:char}(d) correspond to data generated using Model M7. The described characteristics include the proportion of those split points that are within the middle 5\% of a standard normal distribution (i.e., proportion in $I = (-0.063, 0.063)$ since the true split point is zero), the average number of terminal nodes (presented in logarithmic scale) when a split is made and proportion of splits at variables besides $\mathbf{X}_1$ (presented as a percentage).}
    \label{fig:char}
\end{figure}

\begin{table}[ht]
    \centering
    \caption{Type I Error rate when there is no treatment effect heterogeneity (i.e., when $\gamma = 0$ and the data are generated according to Model M1). Probabilities are computed for a tree making a split and variations include the tree splitting at X1 at the root node, splitting at X1 at any node and any split at all. The error rate is determined for TEHTrees (TT), Causal Trees (CT) and Conditional Inference Trees over effects estimated by Causal Forests (CF)}
    \label{tab:subtype}
    \begin{tabular}{|cccccccc|}
    \hline
     Method & N & m & $\rho$ & TN &  Type I (M1) & Type I (M1) & Type I (M1)\\
    & & & & &  $X_1$ - Root node & $X_1$ - Any node & Any split\\
    \hline
      CF  & 100 & 5 & 0.4 & 3.859 & 0.181 & 0.446 & 0.947\\
      TT  & 100 & 5 & 0.4 & 1.034 & 0.004& 0.004& 0.034 \\
     CT & 100 & 5 & 0.4 & 1.404 & 0.077 & 0.086 & 0.43 \\
     
     CF  & 200 & 5 & 0.4 & 7.476 & 0.211 & 0.676 & 0.999 \\
      TT & 200 & 5 & 0.4 & 1.04 & 0.012& 0.012 & 0.036 \\
      CT & 200 & 5 & 0.4 & 2.058 & 0.199& 0.275 & 0.636\\
      
      CF & 500 & 5 & 0.4 & 15.369 & 0.190& 0.92 & 0.998 \\
      TT & 500 & 5 & 0.4 & 1.03 & 0.002& 0.002& 0.028 \\
      CT & 500 & 5 & 0.4 & 4.033 & 0.225& 0.487 & 0.694\\ 
     \hline
     \hline
      CF & 100 & 5 & 0.8 & 4.051 & 0.176& 0.483 & 0.934\\
      TT & 100 & 5 & 0.8 & 1.092 & 0.02& 0.022& 0.078 \\
     CT & 100 & 5 & 0.8 & 1.54 & 0.112 & 0.145 & 0.213 \\
     
      CF  & 200 & 5 & 0.8 & 6.878 & 0.211 & 0.681 & 0.95\\
      TT  & 200 & 5 & 0.8 & 1.068 & 0.018 & 0.02& 0.06\\
      CT & 200 & 5 & 0.8 & 1.987 & 0.145 & 0.213 & 0.601 \\
      
     CF  & 500 & 5 & 0.8 & 14.193& 0.182 & 0.879& 0.961\\
     TT  & 500 & 5 & 0.8 & 1.036 & 0.008 & 0.008 & 0.034\\
     CT & 500 & 5 & 0.8 & 3.29 & 0.168 & 0.364 & 0.667\\ 
     \hline
     \hline
    \end{tabular}
\end{table}

\begin{table}[ht]
    \centering
        \caption{Power for TEHTrees (TT), Causal Trees (CT) and Conditional inference trees over effects estimated by Causal Forests (CF) when data (continuous outcome and five covariates) are generated according to Model M3. Power is defined to be the probability of the tree making a split and variations include the tree splitting X1 at the root node, X1 at any node (including inner nodes) and any split at all (includes any covariate at all levels)}\label{tab:subpower}
    \begin{tabular}{|ccccccccc|}
    \hline
    %0.106 & 0.108  & 0.124
    %.164 & 0.164 & 0.18
    %0.512 & 0.514 & 0.514 
    Method & N & m & $\gamma$& $\rho$ & TN & Power (M3) & Power (M3) & Power (M3) \\
     & & & & &  &X1 - Root node & X1 - Any node & Any split \\
      \hline
      CF  & 100 & 5 & 1 & 0 & 3.771 & 0.565 & 0.774 & 0.989 \\
      TT  & 100 & 5 & 1 & 0 & 1.136 & 0.106 & 0.108 & 0.124 \\
      CT & 100 & 5 & 1 & 0 & 1.389 & 0.255 & 0.233 & 0.384 \\
     
      CF  & 200 & 5 & 1 &0 & 8.313 & 0.819 & 0.979& 1\\
      TT  & 200 & 5 & 1 &0 & 1.206 & 0.172 & 0.172& 0.186 \\
      CT & 200 & 5 & 1 &0 & 2.196 & 0.491 & 0.56 & 0.682 \\
     
      CF  & 500 & 5 & 1 &0 & 18.908 & 0.994 &1 &1\\
     TT & 500 & 5 & 1 &0 & 1.56 & 0.47 & 0.476 & 0.484  \\
     CT & 500 & 5 & 1 &0 & 4.465 & 0.639 & 0.762 & 0.796\\
    
     CF & 1000 & 5 & 1 &0 & 32.21 & 1 & 1 & 1\\
     TT  & 1000 & 5 & 1 &0 & 1.954 & 0.816 & 0.818 & 0.824 \\
     CT & 1000 & 5 & 1 &0 & 8.766 & 0.772 & 0.874 & 0.871\\
    
     CF & 2000 & 5 & 1 &0 & 49.536 & 1 & 1 & 1\\
     TT & 2000 & 5 & 1 &0 & 2.28 & 0.968& 0.976 & 0.98  \\
     CT & 2000 & 5 & 1 &0 & 17.535 & 0.905  & 0.941 & 0.95\\
     \hline
      CF  & 200 & 5 & 1 &0.4 & 9.907 &  0.843 & 0.986& 1 \\
      TT  & 200 & 5 &1 & 0.4 & 1.358 & 0.27 & 0.27 & 0.32 \\
     CT & 200 & 5 & 1 &0.4 & 2.1 & 0.318 & 0.398 & 0.635\\
     
     CF  & 500 & 5 & 1 &0.4 & 21.055 & 0.989 & 1 & 1 \\
     TT  & 500 & 5 & 1 &0.4 & 1.796 & 0.6 & 0.616 & 0.65  \\
     CT & 500 & 5 & 1 &0.4 & 4.12 & 0.417 & 0.634 & 0.74\\ 
     
     \hline
     CF  & 200 & 5 & 1 &0.8 & 10.856 & 0.684 & 0.987 & 0.998\\
     TT  & 200 & 5 & 1 &0.8 & 1.52 & 0.232 & 0.246 & 0.428 \\
     CT & 200 & 5 & 1 &0.8 & 1.979 & 0.202 & 0.268 & 0.63  \\
      
     CF  & 500 & 5 & 1 &0.8 & 22.355 & 0.894 & 1&1 \\
     TT  & 500 & 5 & 1 &0.8 & 1.976 & 0.466 & 0.528 & 0.754\\
     CT & 500 & 5 & 1 &0.8 & 3.415 & 0.257 & 0.438 & 0.706 \\ 
     \hline
     \hline
    \end{tabular}
    
\end{table}

\begin{table}[ht]
    \centering
    \caption{Power for TEHTrees (TT), Causal Trees (CT) and Conditional inference trees over effects estimated by Causal Forests (CF) when data (continuous outcome and five covariates) are generated according to Model M4. Power is defined to be the probability of the tree making a split and variations include the tree splitting X1 at the root node, X1 at any node (including inner nodes) and any split at all (includes any covariate at all levels)}
    \label{tab:subpower2}
    \begin{tabular}{|ccccccccc|}
    \hline
    Method & N & m & $\gamma$ & $\rho$ & TN & Power (M4) & Power (M4) & Power (M4)\\
     & & & & &  & X1 - Root node & X1 - Any node & Any split \\
      \hline
     CF  & 100 & 5 & 1 & 0 & 4.043 & 0.892 & 0.966 & 1\\
     TT  & 100 & 5 & 1 & 0 & 1.862 & 0.698 & 0.7 & 0.704 \\
      CT & 100 & 5 & 1 & 0 & 1.475 & 0.356 & 0.352 & 0.495\\
       CF  & 200 & 5 & 1 & 0 & 9.754 & 0.995 & 1 & 1 \\
    TT  & 200 & 5 & 1 & 0 & 2.616 & 0.966 & 0.966 & 0.966\\
      CT & 200 & 5 & 1 & 0 & 2.621 & 0.813 & 0.844 & 0.872\\
     CF  & 500 & 5 & 1 & 0 & 25.192 & 1 & 1 & 1 \\
     TT  & 500 & 5 & 1 & 0 & 4.252 & 1 &1 &1\\
      CT & 500 & 5 & 1 & 0 & 5.921 & 0.979 & 0.982 & 0.989\\
     \hline
     \hline 
      CF  & 100 & 5 & 2 & 0 & 4.551 & 0.999 & 1 & 1\\
     TT  & 100 & 5 & 2 & 0 & 2.864 & 0.996 & 0.996 & 0.998 \\
      CT & 100 & 5 & 2 & 0 & 1.627& 0.585 & 0.557 & 0.615\\
      CF  & 200 & 5 & 2 & 0 & 10.925 & 1 & 1 & 1\\
      TT  & 200 & 5 & 2 & 0 & 4.012 & 1 & 1& 1\\
      CT & 200 & 5 & 2 & 0 & 2.945 & 0.991 & 0.99 & 0.989 \\
      CF & 500 & 5 & 2 & 0 & 28.615 & 1 & 1 & 1\\
     TT  & 500 & 5 & 2 & 0 & 6.312 & 1&1 &1\\
      CT & 500 & 5 & 2 & 0 & 6.557 & 1 & 1& 1\\
     \hline
    \end{tabular}
\end{table}

\begin{table}[ht]
    \centering
    \caption{Power for TEHTrees (TT), Causal Trees (CT) and Conditional Inference Trees over effects estimated by Causal Forests (CF) when data (continuous outcome and five covariates) are generated according to Model M6. Power is defined to be the probability of the tree making a split and variations include the tree splitting X1 at the root node, X1 at any node (including inner nodes) and any split at all (includes any covariate at all levels)}
    \label{tab:subpower6}
    \begin{tabular}{|ccccccccc|}
    \hline
     Method & N & m & $\gamma$ & $\rho$ & TN & Power (M6) & Power (M6) & Power (M6) \\
    & & & & & &  X1 - Root node & X1 - Any node & Any split \\
      \hline
     CF & 100 & 5 & 1 & 0 & 3.439 & 0.245 & 0.46 & 0.976 \\
     TT & 100 & 5 & 1 & 0 & 1.03 & 0.006 & 0.006 & 0.028 \\
      CT & 100 & 5 & 1 & 0 & 1.298 & 0.116 & 0.109 & 0.315\\
      
      CF  & 200 & 5 & 1 & 0 & 6.809 & 0.384 & 0.764 & 0.998 \\
      TT  & 200 & 5 & 1 & 0 & 1.024 & 0.012 & 0.012 & 0.024 \\
      CT & 200 & 5 & 1 & 0 & 2.175 &  0.357 & 0.399 & 0.634\\
     
      CF  & 500 & 5 & 1 & 0 & 13.218 &  0.488 & 0.828 & 0.983\\
      TT  & 500 & 5 & 1 & 0 & 1.046 & 0.014 & 0.016 & 0.04\\
      CT & 500 & 5 & 1 & 0 & 4.695 & 0.479 & 0.687 & 0.764\\
    
      CF & 1000 & 5 & 1 & 0 & 21.68 & 0.653 & 0.872 & 0.95\\
      TT  & 1000 & 5 & 1 & 0 & 1.152 & 0.044 & 0.044  & 0.106\\
      CT & 1000 & 5 & 1 & 0 & 9.233 & 0.572 & 0.803 & 0.844 \\
     
     CF  & 2000 & 5 & 1 & 0 & 31.426 & 0.642 & 0.796 & 0.904\\
      TT  & 2000 & 5 & 1 & 0 & 1.492 & 0.134 & 0.142 & 0.246\\
      CT & 2000 & 5 & 1 & 0 & 19.588 & 0.704 & 0.925 & 0.932 \\
     \hline
    \end{tabular}
\end{table}

\begin{table}[ht]
    \centering
    \caption{Power for TEHTrees (TT), Causal Trees (CT) and Conditional inference trees over effects estimated by Causal Forests (CF) when data (continuous outcome and five covariates) are generated according to Model M7. Power is defined to be the probability of the tree making a split and variations include the tree splitting X1 at the root node, X1 at any node (including inner nodes) and any split at all (includes any covariate at all levels)}
    \label{tab:subpowerm7}
    \begin{tabular}{|c|cccccc|ccc|}
    \hline
     Method & N & m & $\gamma$ & $\eta$ & $\rho$ & TN & Power (M7) & Power (M7) & Power (M7)  \\
    & & & & & & &  Root node & Any node  & Any split \\
      \hline
       CF & 100 & 5 & 1 & 2 & 0 & 3.968 & 0.636 & 0.825 & 0.993\\
     TT  & 100 & 5 & 1 & 2 & 0 & 1.052 & 0.028 & 0.028 & 0.052 \\
     CT & 100 & 5 & 1 & 2 & 0 & 1.419 & 0.269 & 0.241 & 0.426 \\
      
      CF & 200 & 5 & 1 & 2 & 0 & 8.417 & 0.87 & 0.985 & 1  \\
      TT  & 200 & 5 & 1 & 2 & 0 & 1.046 & 0.036 & 0.038 & 0.044\\
     CT & 200 & 5 & 1 & 2 & 0 & 2.358 & 0.543 & 0.61 & 0.73\\
     
      CF  & 500 & 5 & 1 & 2 & 0 & 13.987 & 0.993 & 0.998 & 1\\
      TT  & 500 & 5 & 1 & 2 & 0 & 1.13 & 0.048 & 0.052 & 0.096\\
      CT & 500 & 5 & 1 & 2 & 0 & 5.278 & 0.745 & 0.853 & 0.891\\
     \hline
      CF  & 100 & 5 & 2 & 2 & 0 & 5.105 & 0.966 & 0.997 & 1\\
      TT & 100 & 5 & 2 & 2 & 0 & 1.068 & 0.048 & 0.05 & 0.06\\
     CT  & 100 & 5 & 2 & 2 & 0 & 1.546 & 0.466 & 0.45 & 0.558\\
      CF  & 500 & 5 & 2 & 2 & 0 & 13.198 & 1 & 1 & 1\\
      TT  & 500 & 5 & 2 & 2 & 0 & 1.294& 0.082 & 0.082 & 0.198\\
      CT  & 500 & 5 & 2 & 2 & 0 & 6.714 & 0.993 & 0.997 & 0.999\\
     \hline
     \hline
      CF  & 100 & 5 & 1 & 1.5 & 0 & 4.075 & 0.762 & 0.904 & 0.997 \\
      TT  & 100 & 5 & 1 & 1.5 & 0 & 1.124 & 0.106& 0.106 &0.12\\
      CT & 100 & 5 & 1 & 1.5 & 0 & 1.441 & 0.303 & 0.279 & 0.471\\
      CF  & 500 & 5 & 1 & 1.5 & 0 & 19.9 & 1 & 1 & 1\\
    TT  & 500 & 5 & 1 & 1.5 & 0 & 1.625 & 0.494 & 0.496 & 0.506 \\
      CT  & 500 & 5 & 1 & 1.5 & 0 & 5.151 & 0.852 & 0.918 & 0.935\\
     \hline
     \hline
      CF & 100 & 5 & 2 & 1.5 & 0 & 4.969 & 0.993 & 1 & 1\\
      TT & 100 & 5 & 2 & 1.5 & 0 & 1.448 & 0.38 & 0.384 &  0.388\\
      CT & 100 & 5 & 2 & 1.5 & 0 & 1.606 & 0.542 & 0.505 & 0.581 \\
      CF & 500 & 5 & 2 & 1.5 & 0 & 19.026 & 1 & 1 & 1\\
      TT & 500 & 5 & 2 & 1.5 & 0 & 2.274 & 0.85 & 0.86 & 0.876\\
      CT & 500 & 5 & 2 & 1.5 & 0 & 5.631 & 0.998 & 1 & 1\\
     \hline
     \hline
    \end{tabular}
\end{table}

\begin{table}[ht]
    \centering
    \caption{Power for TEHTrees (TT), Causal Trees (CT) and Conditional inference trees over effects estimated by Causal Forests (CF) when data (continuous outcome and five covariates) are generated according to Model M5. Power is defined to be the probability of the tree making a split and variations include the tree splitting X1 at the root node, X1 at any node (including inner nodes) and any split at all (includes any covariate at all levels)}
    \label{tab:subpowerm5}
    \begin{tabular}{|c|cccccc|ccc|}
    \hline
    %0.106 & 0.108  & 0.124
    %.164 & 0.164 & 0.18
    %0.512 & 0.514 & 0.514 
    Method & N & m & $\rho$ & $\gamma_1$ & $\gamma_2$ & TN & Power (M5) & Power (M5) & Power (M5) \\
    & & & & & & & Root node & Any node & Any split \\
      \hline
      CF & 200 & 5 & 0 & 1 & 1 & 10.209 & 1 & 1 & 1\\
      TT & 200 & 5 & 0 & 1 & 1 & 2.728 & 1&1 &1\\
      CT & 200 & 5 & 0 & 1 & 1 & 2.758 & 0.942 & 0.945 & 0.946\\
     CF & 500 & 5 & 0 & 1 & 1 & 26.281 & 1 & 1 & 1\\
      TT & 500 & 5 & 0 & 1 & 1 & 3.732 & 1 & 1 & 1 \\
     CT & 500 & 5 & 0 & 1 & 1 & 5.814 & 0.998 & 0.999 & 0.997\\
     \hline
     \hline
      CF & 200 & 5 & 0 & 1 & -1 & 8.781 & 0.87 & 0.992 & 0.999\\
     TT & 200 & 5 & 0 & 1 & -1 & 1.648 & 0.53 & 0.532& 0.546\\
    CT & 200 & 5 & 0 & 1 & -1 & 2.43 & 0.46 & 0.513 & 0.74\\
      CF & 500 & 5 & 0 & 1 & -1 & 21.558 & 0.999 & 1 & 1\\
      TT & 500 & 5 & 0 & 1 & -1 & 2.796 & 0.962 & 0.966 & 0.968 \\
      CT & 500 & 5 & 0 & 1 & -1 & 5.676 & 0.845 & 0.896 & 0.932\\
     \hline
     \hline
     CF & 200 & 5 & 0 & -1 & 1 & 8.812 & 1 & 1 & 1 \\
      TT & 200 & 5 & 0 & -1 & 1 & 1.494 & 0.418 & 0.418 & 0.428\\
     CT & 200 & 5 & 0 & -1 & 1 & 2.257 & 0.45 & 0.513 & 0.658\\
      CF & 500 & 5 & 0 & -1 & 1 & 21.826 & 0.998 & 1 & 1\\
      TT & 500 & 5 & 0 & -1 & 1 & 2.71 & 0.944 & 0.944 & 0.944 \\
     CT & 500 & 5 & 0 & -1 & 1 & 5.204 & 0.746 & 0.864 & 0.879\\
     \hline
     \hline
      CF & 200 & 5 & 0 & -1 & -1 & 10.338 & 1 & 1 & 1 \\
     TT & 200 & 5 & 0 & -1 & -1 & 2.544 & 0.996 & 0.996  & 0.996 \\
     CT & 200 & 5 & 0 & -1 & -1 & 2.864 & 0.999 & 0.998 & 1\\
    CF & 500 & 5 & 0 & -1 & -1 & 26.462 & 1 & 1 & 1\\
     TT & 500 & 5 & 0 & -1 & -1 & 3.066 & 1 & 1 & 1 \\
      CT & 500 & 5 & 0 & -1 & -1 & 6.056 & 1 & 1 & 1\\
     \hline
    \end{tabular}
\end{table}

\begin{table}[ht]
    \centering
    \caption{Power for TEHTrees (TT), Causal Trees (CT) and Conditional inference trees over effects estimated by Causal Forests (CF) when data (continuous outcome and five covariates) are generated according to Model M8. Power is defined to be the probability of the tree making a split and variations include the tree splitting X1 at the root node, X1 at any node (including inner nodes) and any split at all (includes any covariate at all levels)}
    \label{tab:my_label}
    \begin{tabular}{|ccccccc|ccc|}
    \hline
    %0.106 & 0.108  & 0.124
    %.164 & 0.164 & 0.18
    %0.512 & 0.514 & 0.514 
     Method & N & m & $\rho$ & $\gamma_1$ & $\gamma_2$ & TN & Power (M8) & Power (M8) & Power (M8) \\
    & & & & & & & Root node & Any node & Any split \\
     \hline
     CF  & 200 & 5 & 0 & 3 & 3 & 11.264 & 1 & 1 & 1\\
     TT  & 200 & 5 & 0 & 3 & 3 & 3.134 & 0.464  & 0.766 & 0.824\\
     CT & 200 & 5 & 0 & 3 & 3 & 2.676 & 0.473 & 0.715 & 0.868\\
      CF & 500 & 5 & 0 & 3 & 3 & 23.279 & 1 & 1 & 1\\
      TT  & 500 & 5 & 0& 3 & 3 & 4.228 & 0.468 & 0.95 & 0.96 \\
      CT & 500 & 5 & 0& 3 & 3 & 5.929 & 0.498 & 0.944 & 0.983\\
     \hline
     \hline
     CF  & 200 & 5 & 0 & 1 & 1 & 9.529 & 0.931 & 1  & 1\\
    TT  & 200 & 5 & 0 & 1 & 1 & 1.372 & 0.172 & 0.192  & 0.304\\
      CT & 200 & 5 & 0 & 1 & 1 & 2.182 & 0.364 & 0.496 & 0.669\\
     CF  & 500 & 5 & 0 & 1 & 1 & 22.77 & 0.999 & 1 & 1 \\
     TT  & 500 & 5 & 0 & 1 & 1 & 2.178 & 0.342 & 0.516  & 0.682 \\
    CT & 500 & 5 & 0 & 1 & 1 & 5.015 & 0.418 & 0.756 & 0.831\\
     \hline
     \hline
      CF  & 200 & 5 & 0& 3 & -3 & 11.222  & 1 & 1 & 1\\
      TT  & 200 & 5 & 0 & 3 & -3 & 2.428 & 0.31 & 0.556 & 0.88\\
     CT & 200 & 5 & 0 & 3 & -3 & 3.236 & 0.529 & 0.874 & 0.995\\
     CF  & 500 & 5 & 0 & 3 & -3 & 23.295 & 1 & 1 & 1\\
    TT  & 500 & 5 & 0 & 3 & -3 & 3.998 & 0.152 & 0.928 & 1 \\
     CT & 500 & 5 & 0 & 3 & -3 & 6.334 & 0.495 & 0.999 & 1\\
     \hline
     \hline
     CF  & 200 & 5 & 0& 1 & -3 & 10.006 & 1 & 1 & 1\\
     TT  & 200 & 5 & 0 & 1 & -3 & 2.064 & 0.022 & 0.04  & 0.882 \\
     CT & 200 & 5 & 0 & 1 & -3 & 2.889 & 0.002 & 0.426 & 0.996\\
     CF  & 500 & 5 & 0 & 1 & -3 & 21.516  & 1 & 1 & 1\\
     TT  & 500 & 5 & 0 & 1 & -3  & 2.58 & 0 & 0.088 & 1 \\
     CT & 500 & 5 & 0 & 1 & -3 & 5.407 & 0 & 0.761  & 1\\
     \hline
    \end{tabular}
\end{table}

\begin{figure}[ht]
    \centering
    \includegraphics[scale = 0.5]{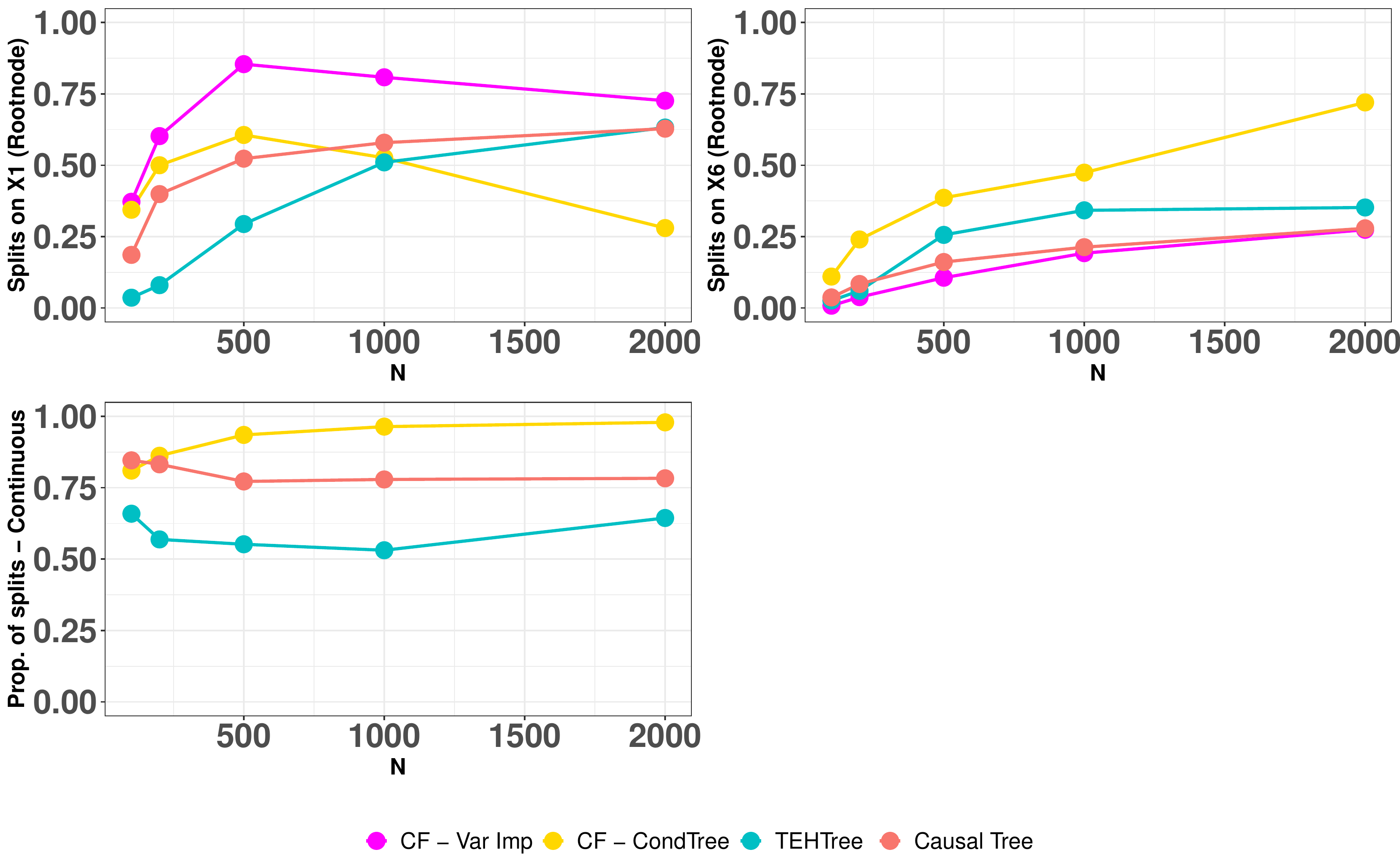}
    \caption{Proportion of splits computed for a scenario with $m=10$ covariates as specified by Model M11. These probabilities are computed over the number of splits on X1 at the rootnode, number of splits on X6 at the rootnode and number of continuous splits.}
    \label{fig:varsel}
\end{figure}

%Figure \ref{fig:varsel} highlights the inherent bias of Causal Trees and related CART-based frameworks towards splitting on covariates that offer many potential splits \citep{strobl2007, song2004study}. We also demonstrate that simply relying on a conditional inference tree framework that uses effect estimates extracted from a Causal Forest may not result in more accurate splits. 

\begin{table}[ht]
    \centering
    \caption{Covariate splits selected using different methods including Causal Tree, TEHTree, Conditional inference tree over effects estimated by a Causal Forest and Importance measure computed for the Causal Forest framework. An equal number of continuous and categorial covariates are fed into this data generating scenario. Characteristics of interest include the proportion of continuous splits, probability of the method splitting on X1 at the rootnode and the probability of the method splitting on X6 at the rootnode.}
    \label{tab:var_selection}
    \begin{tabular}{|c|ccc|ccc|}
    \hline
       Method  &  N & m & $\rho$ & Rootnode & Rootnode & Prop. splits\\
        & & & & X1 & X6 & Continuous\\
        \hline
        CF (I) & 500 & 10 & 0.4  & 0.84 & 0.096 & -\\
        CF & 500 & 10 & 0.4 & 0.706 & 0.27 & 0.942\\
        TT & 500 & 10 & 0.4 & 0.382 & 0.234 & 0.443\\
        CT & 500 & 10 & 0.4 & 0.368 & 0.028 & 0.840 \\ 
        \hline
        CF (I) & 1000 & 10 & 0.4  & 0.8 & 0.198  & -\\
        CF & 1000 & 10 & 0.4 & 0.538 & 0.462 & 0.967\\
        TT & 1000 & 10 & 0.4 & 0.578 & 0.32 & 0.623\\
        CT & 1000 & 10 & 0.4 & 0.403 & 0.057 & 0.810\\ 
        \hline
        CF (I) & 2000 & 10 & 0.4  & 0.73 & 0.27 & -\\
        CF & 2000 & 10 & 0.4 & 0.324 & 0.676 & 0.98\\
        TT & 2000 & 10 & 0.4 & 0.686 & 0.31 & 0.656\\
        CT & 2000 & 10 & 0.4 & 0.508 & 0.096 & 0.789\\ 
        \hline
        CF (I) & 500 & 10 & 0.8  & 0.684 & 0.16 & -\\
        CF & 500 & 10 & 0.8 & 0.644 & 0.192 & 0.946\\
        TT & 500 & 10 & 0.8 & 0.364 & 0.182 & 0.762\\
        CT & 500 & 10 & 0.8  & 0.233 & 0.011 & 0.791\\ 
        \hline
        CF (I) & 1000 & 10 & 0.8  & 0.672 & 0.306  & -\\
        CF & 1000 & 10 & 0.8 & 0.554 & 0.428 & 0.968\\
        TT & 1000 & 10 & 0.8 & 0.556 & 0.254 & 0.714\\
        CT & 1000 & 10 & 0.8  & 0.257 & 0.018 & 0.761\\ 
        \hline
        CF (I) & 2000 & 10 & 0.8 & 0.656 & 0.344 & - \\
        CF & 2000 & 10 & 0.8 & 0.26 & 0.74 & 0.981\\
        TT & 2000 & 10 & 0.8 & 0.674 & 0.252 & 0.651\\
        CT & 2000 & 10 & 0.8 &  0.318 & 0.027 & 0.757 \\
        \hline
    \end{tabular}
\end{table}

\begin{figure}[ht]
\centering
\includegraphics[scale = 0.75]{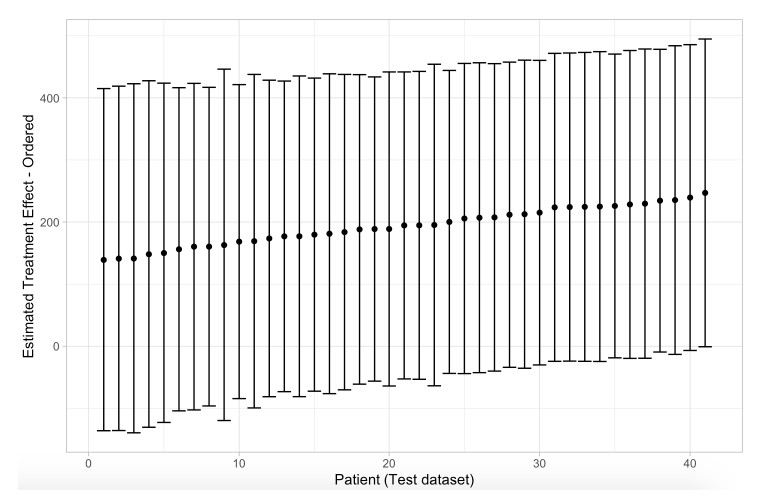}
\caption{Estimated treatment effects (determined using a Causal Forest algorithm) for patients in the test dataset.}\label{fig:plothte}
\end{figure}

\end{document}